\documentclass[%
superscriptaddress,
preprint,
bibnotes,
amsmath,
amssymb,
aps,
pra,
]{revtex4-1}

\usepackage{graphicx}
\usepackage{dcolumn}
\usepackage{bm}
\usepackage{amsxtra}

\usepackage{subcaption}
\usepackage{cleveref}
\begin{document}
\preprint{POP}

\title{Variational Symplectic Particle-in-cell Simulation of Nonlinear Mode Conversion from Extraordinary waves to Bernstein waves}


\author{Jianyuan Xiao}
\affiliation{ Department of Modern Physics and School of Nuclear Science and Technology, University of Science and Technology of China, Hefei, Anhui 230026, China}
\affiliation{Key Laboratory of Geospace Environment, CAS, Hefei, Anhui  230026, China}
\author{Jian Liu}\email{jliuphy@ustc.edu.cn}
\affiliation{ Department of Modern Physics and School of Nuclear Science and Technology, University of Science and Technology of China, Hefei, Anhui 230026, China}
\affiliation{Key Laboratory of Geospace Environment, CAS, Hefei, Anhui  230026, China}
\author{Hong Qin}
\affiliation{ Department of Modern Physics and School of Nuclear Science and Technology, University of Science and Technology of China, Hefei, Anhui 230026, China}
\affiliation{ Plasma Physics Laboratory, Princeton University, Princeton, New Jersey 08543, USA}
\author{Zhi Yu}
\affiliation{Theory and Simulation Division, Institute of Plasma Physics, Chinese Academy of Sciences, Hefei, Anhui 230031, China}

\author{Nong Xiang}
\affiliation{Theory and Simulation Division, Institute of Plasma Physics, Chinese Academy of Sciences, Hefei, Anhui 230031, China}

\begin{abstract}
	In this paper, the nonlinear mode conversion of extraordinary waves in nonuniform magnetized plasmas is studied using the variational symplectic particle-in-cell simulation. The accuracy of the nonlinear simulation is guaranteed by the long-term accuracy and conservativeness of the symplectic algorithm. The spectra of the electromagnetic wave, the evolution of the wave reflectivity, the energy deposition profile, and the parameter-dependent properties of radio-frequency waves during the nonlinear mode conversion are investigated. It is illustrated that nonlinear effects significantly modify the physics of the radio-frequency injection in magnetized plasmas. The evolutions of the radio-frequency wave reflectivity and the energy deposition are observed, as well as the self-interaction of the Bernstein waves and mode excitations. Even for waves with small magnitude, nonlinear effects can also become important after continuous wave injections, which are common in the realistic radio-frequency wave heating and current drive experiments.
\end{abstract}
\maketitle

\newcommand{\EXP}{\times 10^}
\newcommand{\rmd}{\mathrm{d}}
\newcommand{\xs}{ \mathbf{x}_s}
\newcommand{\dotxs}{\dot{\mathbf{x}}_s}
\newcommand{\bfx}{\mathbf{x}}
\newcommand{\bfv}{\mathbf{v}}
\newcommand{\bfA}{\mathbf{A}}
\newcommand{\bfB}{\mathbf{B}}
\newcommand{\bfE}{\mathbf{E}}
\newcommand{\bfu}{\mathbf{u}}
\newcommand{\bfe}{\mathbf{e}}
\newcommand{\rme}{\mathrm{e}}
\newcommand{\rmi}{\mathrm{i}}
\newcommand{\rmq}{\mathrm{q}}
\newcommand{\ope}{\omega_{pe}}
\newcommand{\oce}{\omega_{ce}}
\newcommand{\FIG}[1]{Fig.~\ref{#1}}
\newcommand{\EQ}[1]{Eq.~(\ref{#1})}
\newcommand{\SEC}[1]{Sec.~\ref{#1}}
\newcommand{\REF}[1]{Ref.~\cite{#1}}
\newcommand{\cpt}{\captionsetup{justification=raggedright
}}

\newcommand{\act}{\mathcal{A}}
\section{Introduction}
Radio-frequency (rf) wave injection plays an important role in auxiliary heating and current drive for modern magnetic confinement fusion devices \cite{ushigusa1989lower,lianmin20102450,start1998dt,laqua1997resonant}. A typical method for rf wave injection into overdense plasmas, where the plasma frequency $\omega_{pe}$ is much larger than the electron cyclotron frequency $\omega_{ce}$, is to excite the electron Bernstein waves (EBWs). Two different approaches are commonly used for coupling rf power to EBWs, i.e., the ordinary-extraordinary-Bernstein (O-X-B) and extraordinary-Bernstein (X-B) mode conversion. The O-X-B scheme is usually used for the electron Bernstein wave current drive (EBCD) in spherical tokamaks (STs), where the electron cyclotron resonance current drive (ECCD) method is ineffective \cite{cairns2000prospects,laqua1997resonant}. The X-B scheme requires steeper plasma density profile, which is also available in some STs \cite{kuehl1967coupling,ram2000excitation,jones2003controlled,shiraiwa2006heating}. During the X-B mode conversion, the electromagnetic wave is initially injected into the plasma perpendicularly to the background magnetic field, and propagates as an extraordinary (X) wave within the plasma. When the X wave reaches the cutoff location, where the local right-handed cutoff frequency $\omega_R$ is equal to the frequency of the incident wave $\omega$, part of the energy of the incident wave penetrates the cutoff layer and continues propagating as EBWs. The X-B conversion will be significantly enhanced if the electron density gradient becomes large enough.

The linear X-B mode conversion can be theoretically investigated using the cold plasma model \cite{budden1988propagation,ram1996mode}. In this model, the kinetic effects of the electrons are neglected. At the same time, the ions offer a positive charged background. The brief description of the linear theory is presented in Appendix \ref{SecLinearTheory}. The linear theory predicts that the reflectivity of the injected wave is a constant that does not change with time.

The cold plasma assumption is only accurate when the wave length is much larger than the electron gyroradius. As the electron temperature increases, the kinetic effect becomes significant and should be taken into account. The kinetic theory of mode conversion is so complicated that the analytical solution is unaccessible except for some special cases \cite{kuehl1967coupling}. In addition, the linear condition no longer holds for large-amplitude or long-term wave evolution. In modern magnetic confinement devices, the power of the rf wave injection is so large that the nonlinear effects can not be neglected. The nonlinear effects during the propagation of Bernstein waves have been directly observed in some experiments \cite{sugawa1988observation,porkolab1985nonlinear,laqua1997resonant}. Those nonlinear effects can modify the reflectivity during the wave injection, which cannot be treated by the linear theory. The simulation study on the nonlinear physics of rf waves in plasmas has become an active research field \cite{xiang2006low,yu2009gyrocenter,liunonlinear,PhysRevLett.100.085002}. However, the analytical and numerical studies on nonlinear rf physics are severely constrained by existing mathematical and algorithmic tools.

Containing both kinetic and nonlinear effects, first-principle particle simulation is regarded as a necessary and powerful tool for dealing with the nonlinear rf physics. Different aspects of the X-B mode conversion have lately been investigated using particle-in-cell (PIC) codes \cite{xiang2006low,yu2009gyrocenter,liunonlinear,PhysRevLett.100.085002}. For nonlinear mode conversions, where the timescale of the mode conversion is much larger than the periods of the rf waves, long-term simulations are inevitably required. Since the number of time steps in a long-term simulation is large, the coherent accumulation of the numerical error from each time step turns out to be the most serious difficulty. To ensure the reliability of the simulation results, long-term numerical accuracy and stability should be guaranteed. To overcome the difficulty, a newly developed variational symplectic PIC method \cite{xiao2013variational} is employed in this paper to study the nonlinear X-B mode conversion. The symplectic PIC algorithm bounds the global numerical error for an arbitrary long time by preserving the discrete symplectic structure of the original Lagrangian system. Its long-term accuracy, stability, and conservativeness have been theoretically discussed and practically verified \cite{hairer2006geometric,xiao2013variational}.

Suffering no accumulation of the global errors, the symplectic PIC method shows its unparalleled advantage in the long-term simulations and provides accurate nonlinear results. In the simulation of nonlinear rf mode conversion, it is illustrated that the nonlinear effects significantly modify the physics of the rf wave injection in magnetized plasmas. The wave reflectivity, which is regarded as a constant in linear theory, varies in the nonlinear mode conversion process. The evolutions of the rf wave reflectivity and the energy deposition are observed, as well as the self-interaction of the Bernstein waves and mode excitations. The spectra of the electromagnetic wave, the energy deposition profile, and the parameter-dependent properties of radio-frequency waves during the nonlinear mode conversion are also investigated. It is found that even for waves weak in magnitude, nonlinear effects can become significant after continuous wave injections. When simulating rf waves with small amplitudes, nonlinear phenomena become prominent after several hundred wave periods. New modes are excited over a sufficient long time even if the amplitude of incident wave is small. To study the parameter dependency of the nonlinear X-B conversion, we compare the rf wave injection processes with different electron temperatures, amplitudes of injection waves, and plasma density gradients. It is discovered that a lower electron temperature can enhance the reflectivity, and the reflectivity depends on the plasma density gradient in a sensitive way.

The paper is organized as follows. In Sec.~\ref{SecSympTheory}, we introduce the physical model and the problem of the nonlinear X-B mode conversion. The numerical method used in the long-term nonlinear simulation, i.e., the variational symplectic PIC algorithm, is also introduced. In Sec.~\ref{SecResults}, the symplectic PIC simulation of the nonlinear X-B conversion is carried out. Different factors, which effect the nonlinear rf wave processes, are analyzed in detail. Section \ref{SecDiscuss} is a the brief summary and future research plan.

\section{The model of X-B conversion and the symplectic PIC method}\label{SecSympTheory}

In the X-B conversion process, the incident wave is perpendicular to the background magnetic field. This problem can thus be described by a planar model. Consider a planar plasma with the density nonuniformity in the $x$-direction and a uniform background magnetic field along the $z$-direction. The wave source is located outside the plasma. The source of the electromagnetic wave, with the electric field component along the $y$-direction, excites an X wave in the plasma boundary. The X wave propagates in the plasma until it reaches a cut-off location, and part of the wave converts to
the Bernstein wave. There are two main restrictions in the
PIC simulation of the X-B mode conversion.
Firstly, the wave length of the Bernstein wave is
much shorter than the electromagnetic wave with the same frequency
in the vacuum. So the grid length in the simulation should
be small enough to resolve it, which means the time step is restricted
to a small value correspondingly. A large time
step will lead to numerical instability because the particles can
pass through multiple grids within a single time step.
Another issue is that the group velocity of the Bernstein wave is
much slower than the speed of light in the vacuum, so
the mode conversion process lasts for a long time. It takes
hundreds of periods before the wave reaches
the core region of plasma.
Then at least ten thousand
time steps are required to represent a full X-B mode conversion process.
Traditional numerical methods may fail to obtain useful results due to
the accumulation of numerical errors.
In order to study the mode conversion process with more accuracy, 
a variational
symplectic method, which has excellent long-term conservation properties \cite{hairer2006geometric},
is adopted in the particle-in-cell simulation.

The basic idea of constructing a variational symplectic
scheme is to discretize the Lagrangian and perform the
discrete variational principle \cite{marsden2001discrete}.
The starting point
is to consider the self-consistent system consisting of
charged particles and electromagnetic fields.
The Lagrangian is the integral
\begin{eqnarray}\label{EqnLagS}
	L\left( \bfx_s,\bfA,\phi;t \right)&=& \iiint_\Omega \mathcal{L}\rmd^3 \bfx~,
\end{eqnarray}
where
\begin{align}\label{EqnLagDensity}
	\mathcal{L}=&\frac{\epsilon_0}{2}\left[-\dot\bfA\left( \bfx ,t\right)-\nabla \phi\left( \bfx ,t\right)\right]^2-\frac{1}{2\mu_0}\left[\nabla\times \bfA\left( \bfx,t \right)\right]^2 \notag\\
&+\sum_{s=1}^{N} \delta\left( \bfx-\bfx_s \right)\left[\frac{1}{2}m_s\dot{\bfx}_s^2+q_s\bfA\left( \bfx ,t\right)\cdot\dot{\bfx}_s - q_s\phi\left( \bfx,t \right)\right]
\end{align}
is the Lagrangian density.
Here $m_s$, $q_s$, $\bfx_s$, and $\dot{\bfx}_s$
denote the mass, electric charge, velocity, and
position of the $s$-th particle respectively, $\epsilon_0$
and $\mu_0$ are the permittivity and
permeability of free space, and $\bfA$ and $\phi$ are
vector potential and scalar potential of the electromagnetic
fields. The time derivative of vector potential is denoted as $\dot{\bfA}$.

We follow the method used in Ref.~\cite{xiao2013variational} to discretize
the Lagrangian $L$ presented in \EQ{EqnLagS} in cubic meshes and perform the discrete variational
principle to obtain the numeric scheme. The discrete Lagrangian is
\begin{equation}\label{EqnLDDEF}
	L_d(l,l+1)=L_{f}+L_{I}+L_{p}~,
\end{equation}
where
\begin{eqnarray}
	L_{f} &=&  \iiint_\Omega \frac{1}{2}\left(\epsilon_0\left( \dot{\bfA}\left( \bfx ,(l+\frac{1}{2})\Delta t\right) \right)^2-\frac{1}{\mu_0}\left( \nabla\times\bfA\left( \bfx ,(l+\frac{1}{2})\Delta t\right) \right)^2\right)\rmd \bfx~,\\
	L_{p} &=&   \iiint_\Omega\sum_s\frac{1}{2}S(\bfx_s-\bfx,a)m_s\dot{\bfx}_{s,l}^2\rmd\bfx=\sum_s\frac{1}{2}m_s\dot{\bfx}_{s,l}^2~,\\
L_{I}&=&  \iiint_\Omega\sum_sq_sS(\bfx_{s,l}-\bfx,a)\left( \bfA\left( \bfx ,l\Delta t\right)\cdot\dot{\bfx}_{s,l} \right)\rmd \bfx~,
\end{eqnarray}
\begin{eqnarray}
\bfA\left( \bfx , (l+\frac{1}{2})\Delta t \right)&=& \sum_{i,j,k} \frac{1}{2}\left( \bfA_{i,j,k,l}+\bfA_{i,j,k,l+1} \right) W(\bfx-\bfx_{i,j,k})~,\\
\bfA\left( \bfx , l\Delta t \right)&=& \sum_{i,j,k} \bfA_{i,j,k,l} W(\bfx-\bfx_{i,j,k})~,\\
\dot\bfA\left( \bfx ,(l+\frac{1}{2})\Delta t\right)&=& \sum_{i,j,k} \left( \bfA_{i,j,k,l+1} - \bfA_{i,j,k,l} \right) W(\bfx-\bfx_{i,j,k})/\Delta t~,\\
\dot\bfx_{s,l}&=& (\bfx_{s,l+1}-\bfx_{s,l})/\Delta t~,
\end{eqnarray}
\begin{eqnarray}
	W(\bfx)=\left\{
	\begin{array}{lc}
		(1-|x|/\Delta x)(1-|y|/\Delta x)(1-|z|/\Delta x)~, & \,\,\,\,\,\,\,\,\, |x|, |y|, |z|<\Delta x~,\\
		0~,& \,\,\,\,\,\,\,\,\, \mathrm{otherwise}~.
	\end{array}
	\right.
\end{eqnarray}
Here $\Delta x$ and $\Delta t$ are the grid size and time step respectively. The subscript $l$ stands for variables in the $l$th time interval, and
$i,j,k$ are integer indices for three directions of the 
Cartesian coordinate frame. The electric potential $\phi$ is set to be zero because of the adoption of temporal gauge.
The corresponding discrete action integral $\mathcal{A}_d$ is
\begin{eqnarray}
	\mathcal{A}_d=\sum_l \Delta t L_d(l,l+1)~,
\end{eqnarray}
and the motion equations are obtained from the discrete variational principle
\begin{eqnarray}
\label{EqnDMEA}\frac{\partial {\act_d}}{\partial \bfA_{i,j,k,l}}&=& 0~,\\
\label{EqnDLF}\frac{\partial {\act_d}}{\partial \bfx_{s,l}}&=& 0~.
\end{eqnarray}

The discretization of Lagrangian and the
discrete variational principle lead to a PIC iteration scheme.
Theoretically, the variational symplectic PIC scheme automatically conserves a discrete symplectic
two-form \cite{marsden2001discrete,qin2009variational}.
For any discrete Lagrangian $L_d (q_l,q_{l+1})$, where $q_l$ denotes the generalized coordinates discretized at the $l$’th time step, there exists an discrete symplectic two-form defined as 
\begin{eqnarray}\label{EqnDTwoForm}
	\Omega_{l,l+1}&=& \frac{\partial L_d(q_l,q_{l+1})}{\partial q_l\partial q_{l+1}}\rmd q_l\wedge q_{l+1}~,
\end{eqnarray}
where the operator d denotes the exterior derivative. 
The discrete Lagrangian of Vlasov-Maxwell system is defined in \EQ{EqnLDDEF} with $q_l=(A_{i,j,k,l},x_{s,l} )$. The two-form in \EQ{EqnDTwoForm} 
is a geometric structure, called the discrete symplectic structure. 
Because the original self-consistent Vlasov-Maxwell system is Hamiltonian, 
it is naturally equipped with a continuous symplectic structure, which is 
conserved exactly during the time evolution. The conserved discrete symplectic structure 
in the variational symplectic PIC method is an approximation of the original
continuous geometric structure. The good long-term properties 
of the symplectic algorithm are guaranteed through the conservation of the discrete symplectic
two-form, that is $\Omega_{l-1,l}=\Omega_{l,l+1}$ for any $l$, 
during time advance in the numerical simulation, which can be proved exactly \cite{marsden2001discrete}. 
Therefore, the numerical method has long-term conservativeness and accuracy, 
which are necessary when simulating long-term nonlinear rf wave problems and have been verified numerically \cite{xiao2013variational}.

To decrease the numerical noise caused by the lack of sampling points, the shape function
is applied.
In this paper it is chosen as
\begin{equation}\label{EqnSmo3}
	S(\bfx;a)=S_1(x;a)S_1(y;a)S_1(z;a)~,
\end{equation}
where
\begin{equation}\label{EqnSmo1}
	S_1(x;a)=\alpha_1\left\{
	\begin{array}{lc}
		1,&\,\,\,\,\,\,\,\, |x|\leq a~,\\
		0,&\,\,\,\,\,\,\,\, |x|>a~,
	\end{array}
	\right.
\end{equation}
and $\alpha_1=1/(2a)$. The factor $a$ is set to be $0.25\Delta x$.

\begin{figure}
	\cpt
\includegraphics[width=0.5\linewidth]{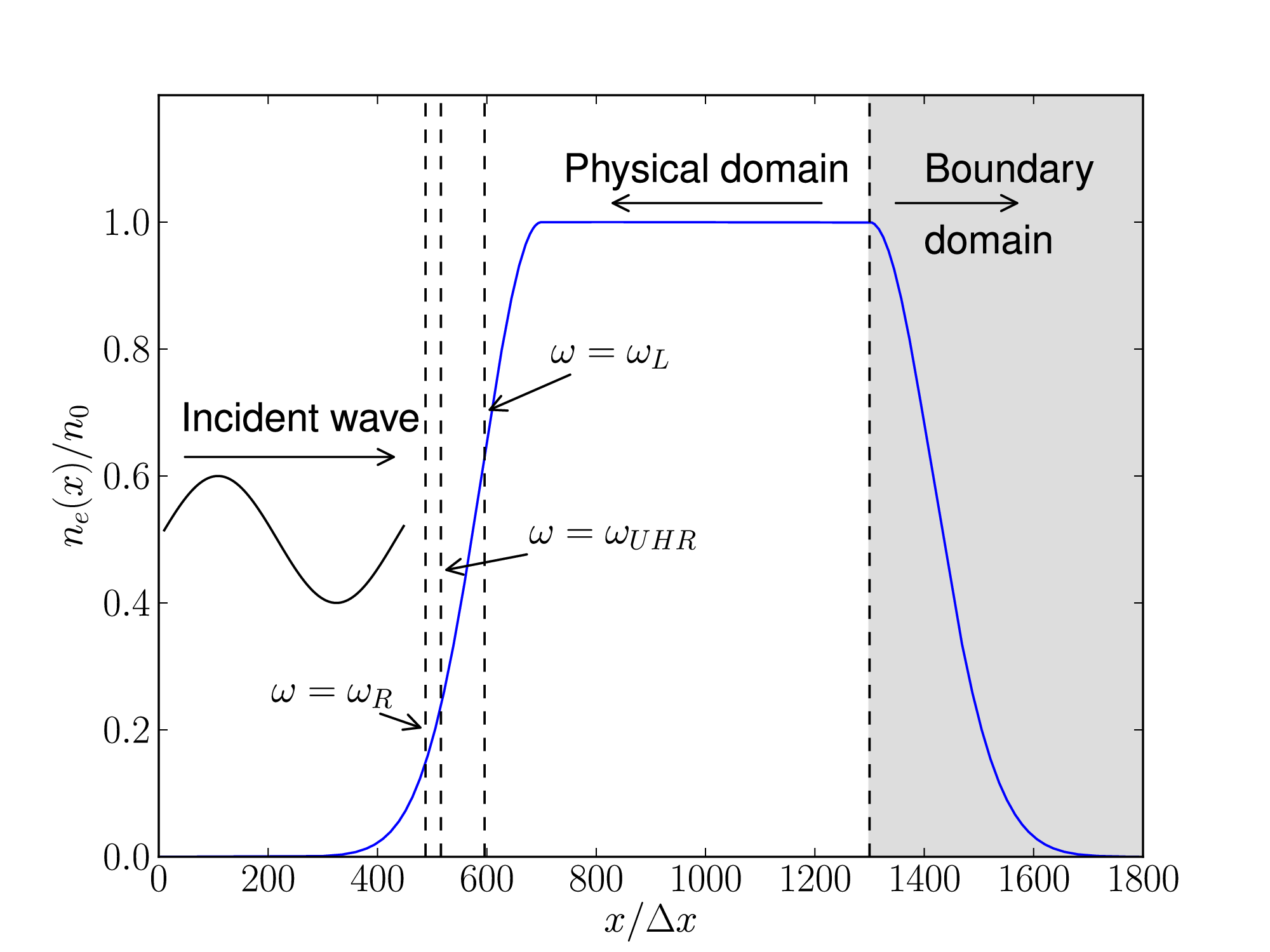}
	\caption{Density profile in X-B conversion simulation examples. The $\omega_R$ and $\omega_L$ are the right and left cutoff frequencies for X wave. Here $n_r=380$ and $\omega=1.62\omega_{ce}$.}
	\label{FigDensProf}
\end{figure}

With the variational symplectic PIC method, we are able to simulate the nonlinear process of 
X-B mode conversion.
The plasma is inhomogeneous in $x$
direction, and as an example we assume the electron density profile is
\begin{eqnarray}\label{EqnDenExample1}
	n_e(x)=n_0\left\{
	\begin{array}{lc}
		\exp\left( -(x/\Delta x-(n_r+320))^2/(2(n_r/3.5)^2) \right)~,&0<x\leq \left( n_r+320 \right) \Delta x~,\\
		1~, &\left( n_r+320 \right),\Delta x<x\leq 1300\Delta x~,
	\end{array}
	\right.
\end{eqnarray}
where $n_r\Delta x$ is the thickness of the plasma boundary,
$\Delta x=2.773\EXP{-5}$~m,
and $n_0=2.0\EXP{19}~\mathrm{m}^{-3}$. The density profile is plotted
in \FIG{FigDensProf}.
The physical domain is $0<x<1300\Delta x$. At $x= 0$ 
we use the Mur's absorbing
boundary condition (MABC) \cite{mur1981absorbing} to 
eliminate the unwanted wave
reflections. However at $x=1300\Delta x$ the
density of plasma is not zero and the MABC fails.
To construct an absorbing boundary, we apply an auxiliary 
calculation domain 1300$\Delta x<x<1792\Delta x$ with the decreasing 
electron density
\begin{eqnarray}
	n_e(x)=n_0 \exp\left( -(x/\Delta x-1300)^2/26122 \right)~.
\end{eqnarray}
During the simulation, the Bernstein wave propagates in the plasma but not fast enough to reach the boundary position
$x=1300\Delta x$.
At $x=1792\Delta x$ the MABC is adopted. The full simulation domain
is the $1792\times 2 \times 2$ cubic mesh, and periodic boundary
conditions are adopted in both $y$ and $z$-directions.
The simulation domain is small (about $4\EXP{-2}$~m) compared with
the fusion device, so the external static magnetic field can be treated as
a constant. We set $\bfB_0=B_0\bfe{_z}$
with $B_0=0.55\mathrm{T}$.
The size of the plasma boundary and the electron temperature 
are set according to 
experimental data \cite{maingi2003h,PhysRevLett.75.4421}
in National Spherical Torus Experiment (NSTX),
where the $n_e/\nabla n_e$ is about $0.01$~m
and the electron temperature $T_e$ is about $100$~eV.
We set $n_r$ in Eq.~(\ref{EqnDenExample1}) to be $380$
in all examples except for cases stated otherwise.
The electron temperature
$T_e$ is chosen to be different constant values independent of spatial positions in different problems,
and the distribution of electron velocity is set to be Maxwellian.
The wave source is located at $x=4\Delta x$. It excites a
linearly polarized
electromagnetic field
as
\begin{eqnarray}
	\bfE(4\Delta x,t)=E_1\mathbf{e}_y\sin\left(\omega t  \right)~,
\end{eqnarray}
where $E_1$ is the amplitude of the oscillating electric field, and $\omega$ is the frequency of incident wave.
The time step is set to $\Delta t=\Delta x/\mathrm{c}\approx 9.24\EXP{-14}$~s.
According parameters, the corresponding electron cyclotron
frequency is $\omega_{ce}=9.67\EXP{10}$~rad/s, and the plasma frequency
is $\omega_{pe_0}=25.2\EXP{10}$~rad/s in the core plasma.
The frequency of the incidence wave $\omega$ is chosen to be
$\omega=0.0145/\Delta t\approx 1.62\omega_{ce}$.
According to the linear mode conversion theory for
extraordinary waves (see Appendix \ref{SecLinearTheory}),
the wave absorption efficiency is high with the above parameters. To suppress the numerical noise in the full-f PIC simulation, the number of sampling points should be as large as possible. However, the number of sampling points is limited by the computing power available. To satisfy the constraints of both the numerical accuracy and the computing power, the number of sampling points per grid at the core plasma is set to be 4000 in the numerical cases. With these methods and parameters, the nonlinear X-B process is numerically studied the next section.

\section{Simulations of X-B mode conversion}\label{SecResults}
\subsection{Full wave solution in nonlinear X-B mode conversion}
\begin{figure}
	\cpt
	\begin{center}
		\includegraphics[width=0.55\linewidth]{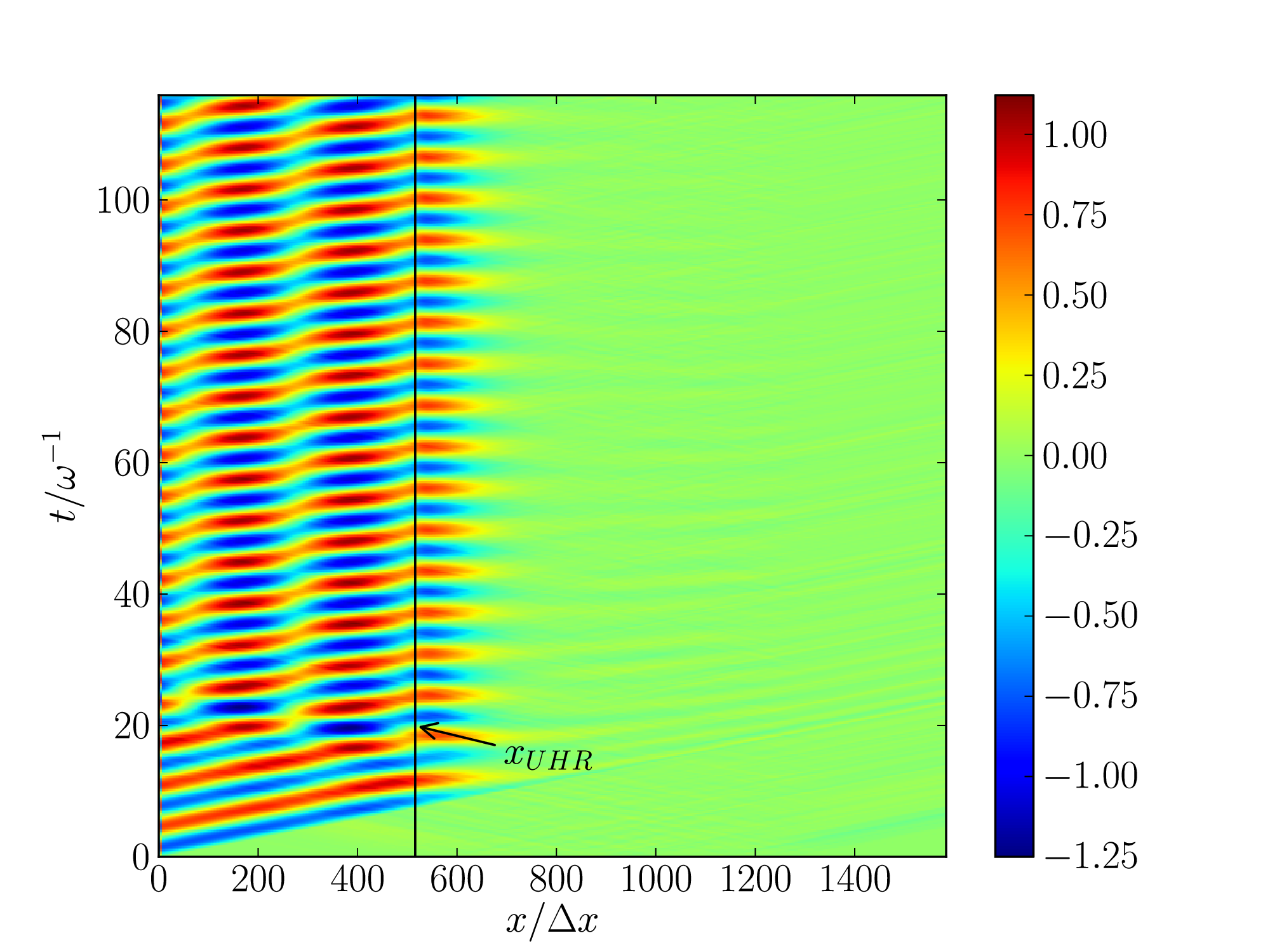}
	\end{center}
	\caption{The contour of $E_y(x,t)$ in the time interval $[0,115/\omega]$}
	\label{FigDen95Temp15Elec0}
\end{figure}

\begin{figure}
	\cpt
	\begin{center}
		\includegraphics[width=0.55\linewidth]{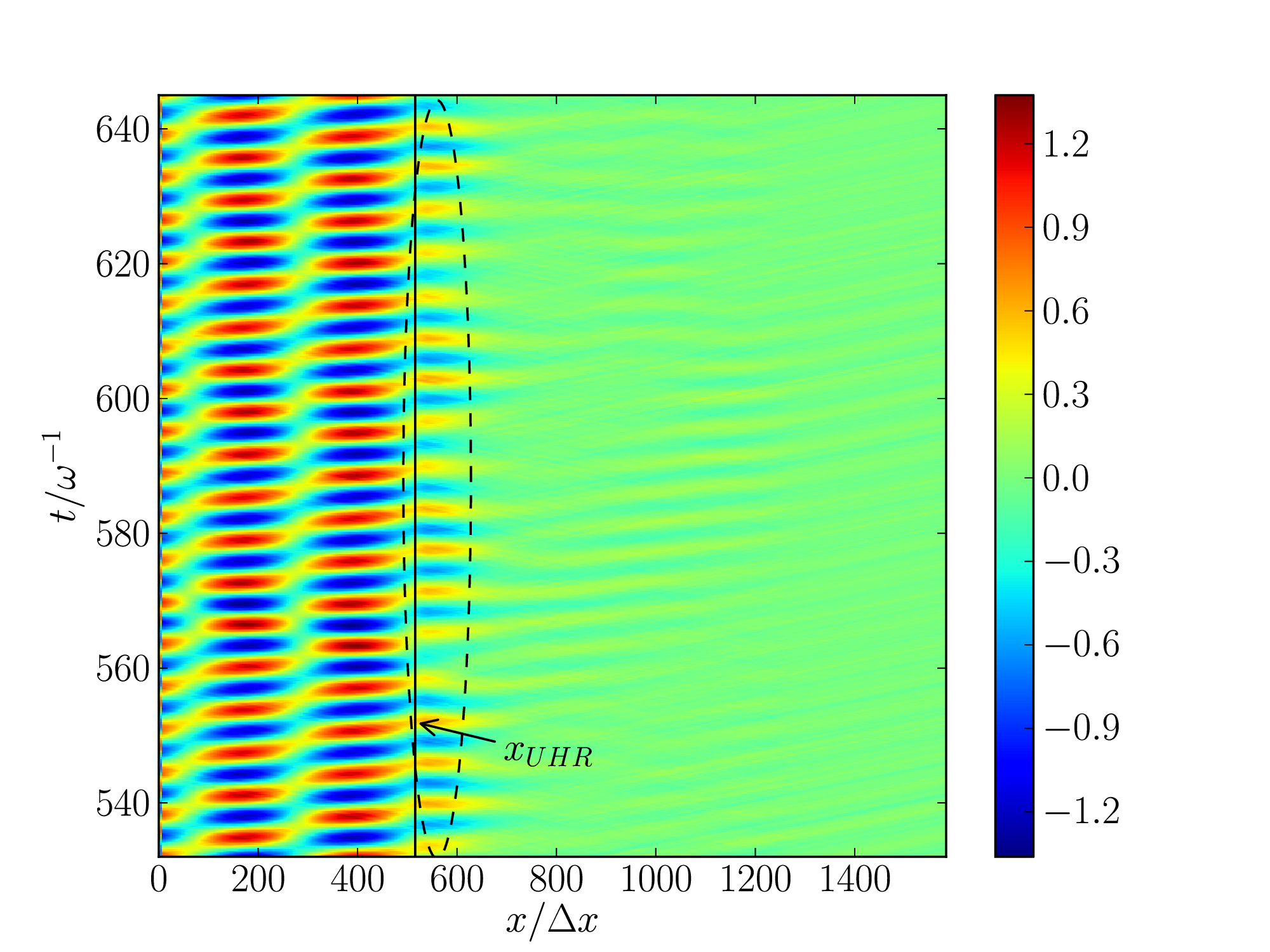}
	\end{center}
	\caption{The contour of $E_y(x,t)$ in the time interval $[530/\omega,645/\omega]$}
	\label{FigDen95Temp15Elec}
\end{figure}

The nonlinear X-B mode conversion can be clearly demonstrated by
the evolution of $E_y(x,t)$. The amplitude of
the wave source $E_1$ is set to $1$~MV/m,
and the temperature of electrons is set to 57.6~eV. 
The evolution of $E_y$
in different time intervals are recorded as contour plots
in \FIG{FigDen95Temp15Elec0} and \FIG{FigDen95Temp15Elec}, where
the horizontal axis is the location $x$ and the vertical axis is the time $t$.
At the beginning of the wave injection, the mode conversion structure forms in a short time, and 
the reflection is stable (\FIG{FigDen95Temp15Elec0}). After about 100 periods, the wave structure to the right of UHR becomes unstable, as marked by the dashed ellipse in  \FIG{FigDen95Temp15Elec}. The wave structure varies with time irregularly, which is an implication of the nonlinear effect. There is no direct evidence of nonlinear generation of other frequencies in Fig. 3. Further results of nonlinear generation of EBWs with other frequencies will be discussed in \SEC{SecSPECTRUMS}.

\subsection{New mode generation during X-B mode conversion} \label{SecSPECTRUMS}
\begin{figure}
	\cpt
	\begin{center}
		\begin{subfigure}[b]{0.42\textwidth}
			\includegraphics[width=\textwidth]{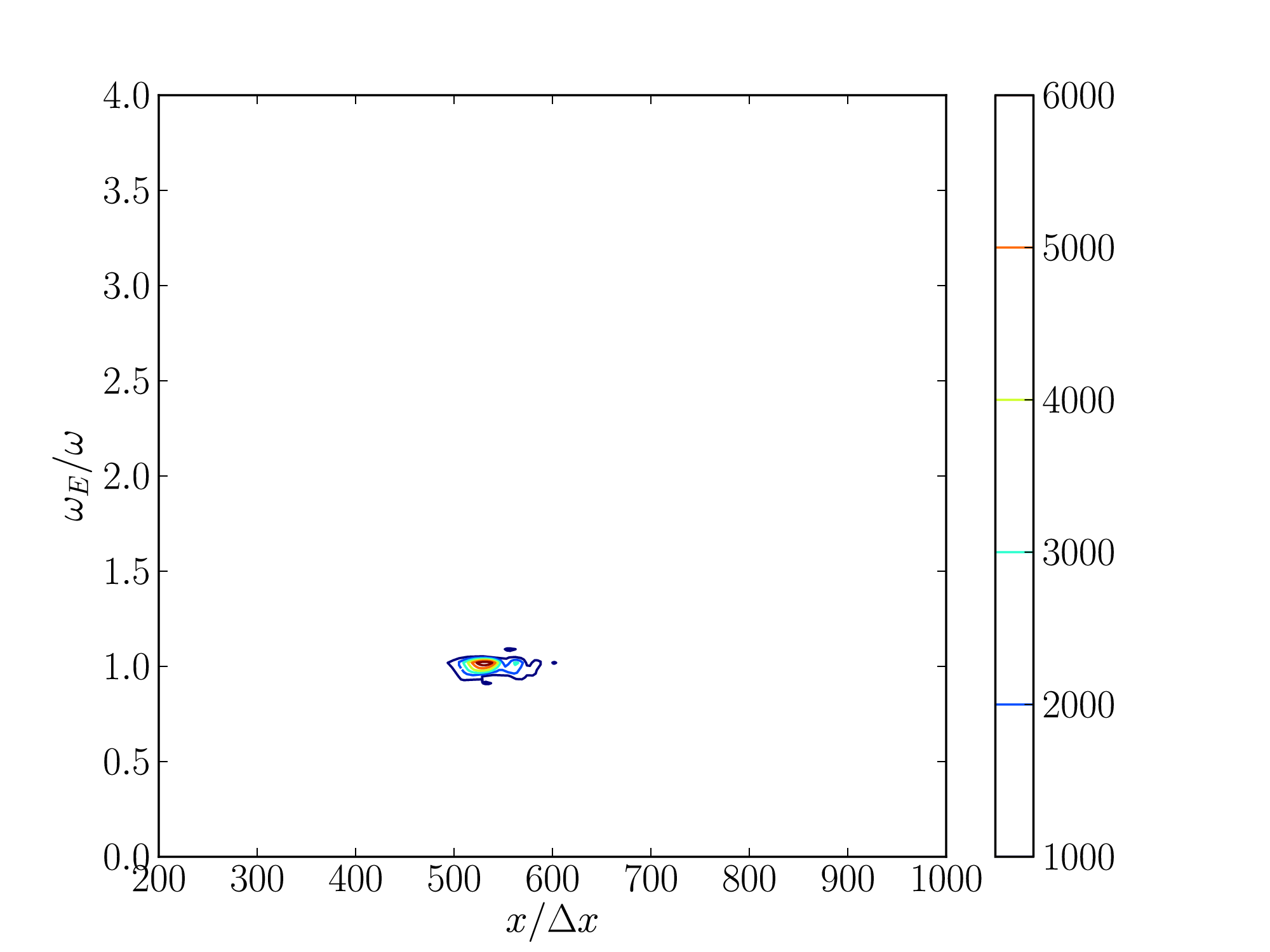}
			\caption{$[0,174/\omega]$}
			\label{FigSpecEx00}
		\end{subfigure}
		\begin{subfigure}[b]{0.42\textwidth}
			\includegraphics[width=\textwidth]{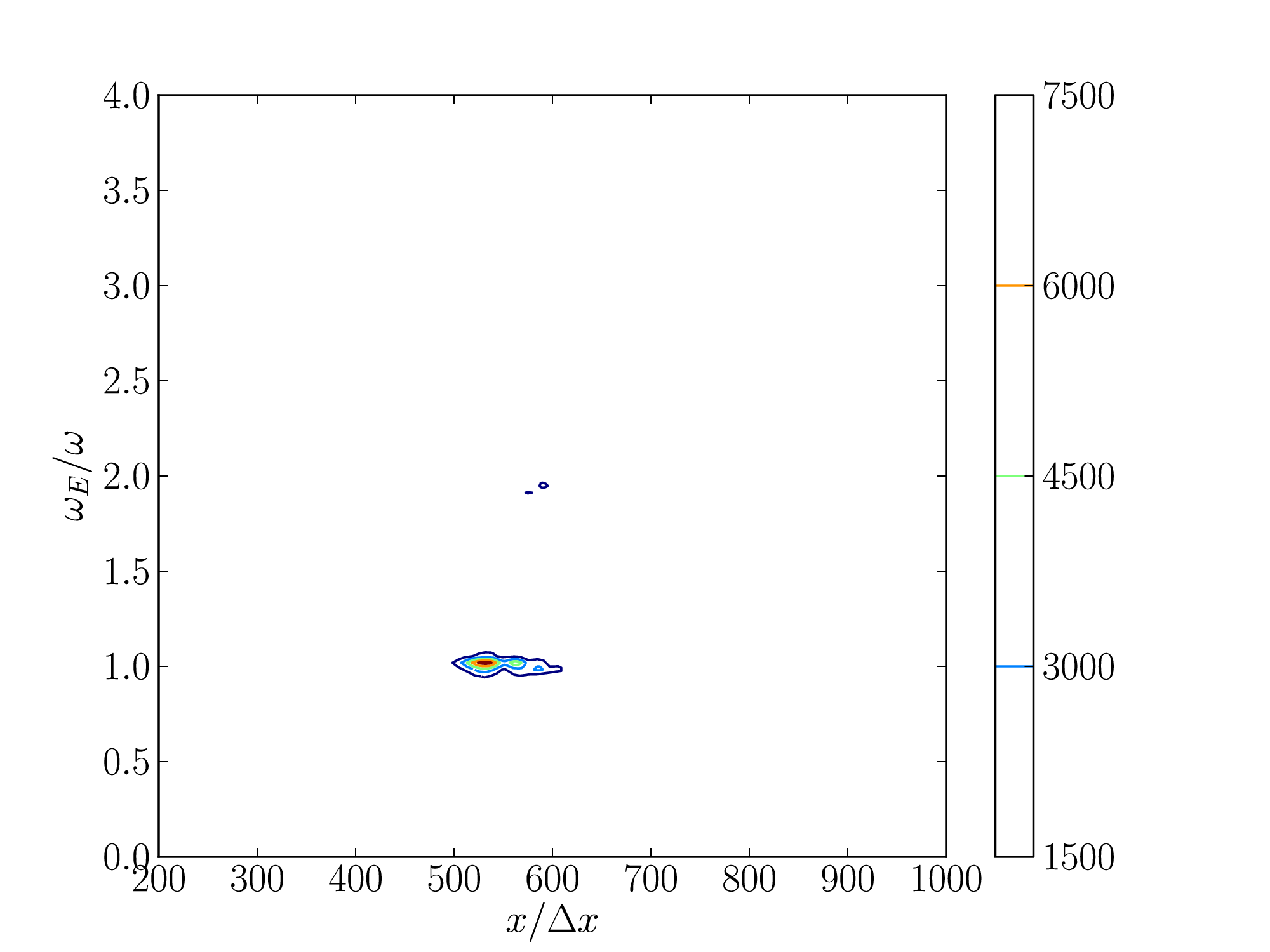}
			\caption{$[116/\omega,290/\omega]$}
			\label{FigSpecEx02}
		\end{subfigure}
		\begin{subfigure}[b]{0.42\textwidth}
			\includegraphics[width=\textwidth]{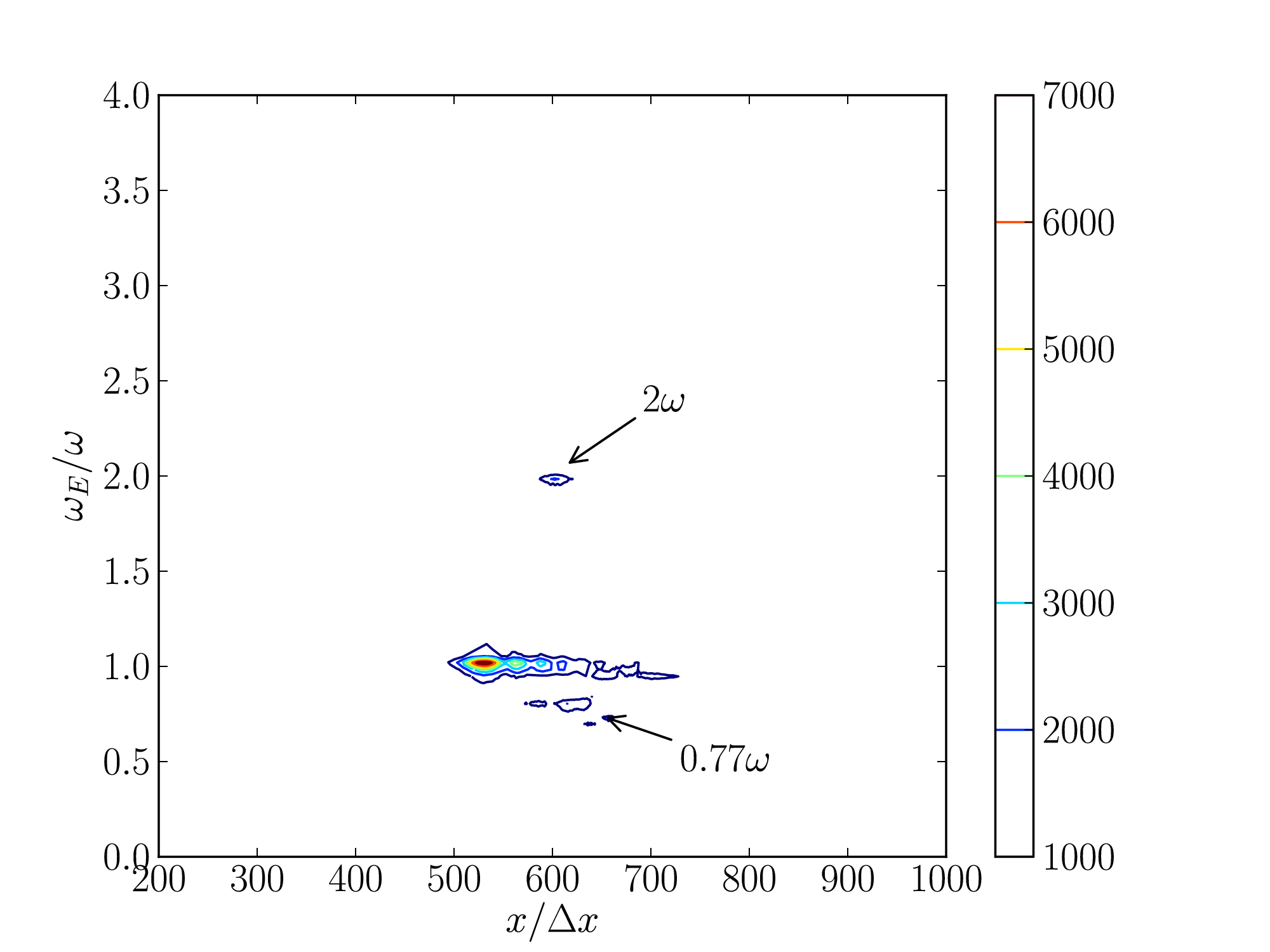}
			\caption{$[290/\omega,464/\omega]$}
			\label{FigSpecEx05}
		\end{subfigure}
		\begin{subfigure}[b]{0.42\textwidth}
			\includegraphics[width=\textwidth]{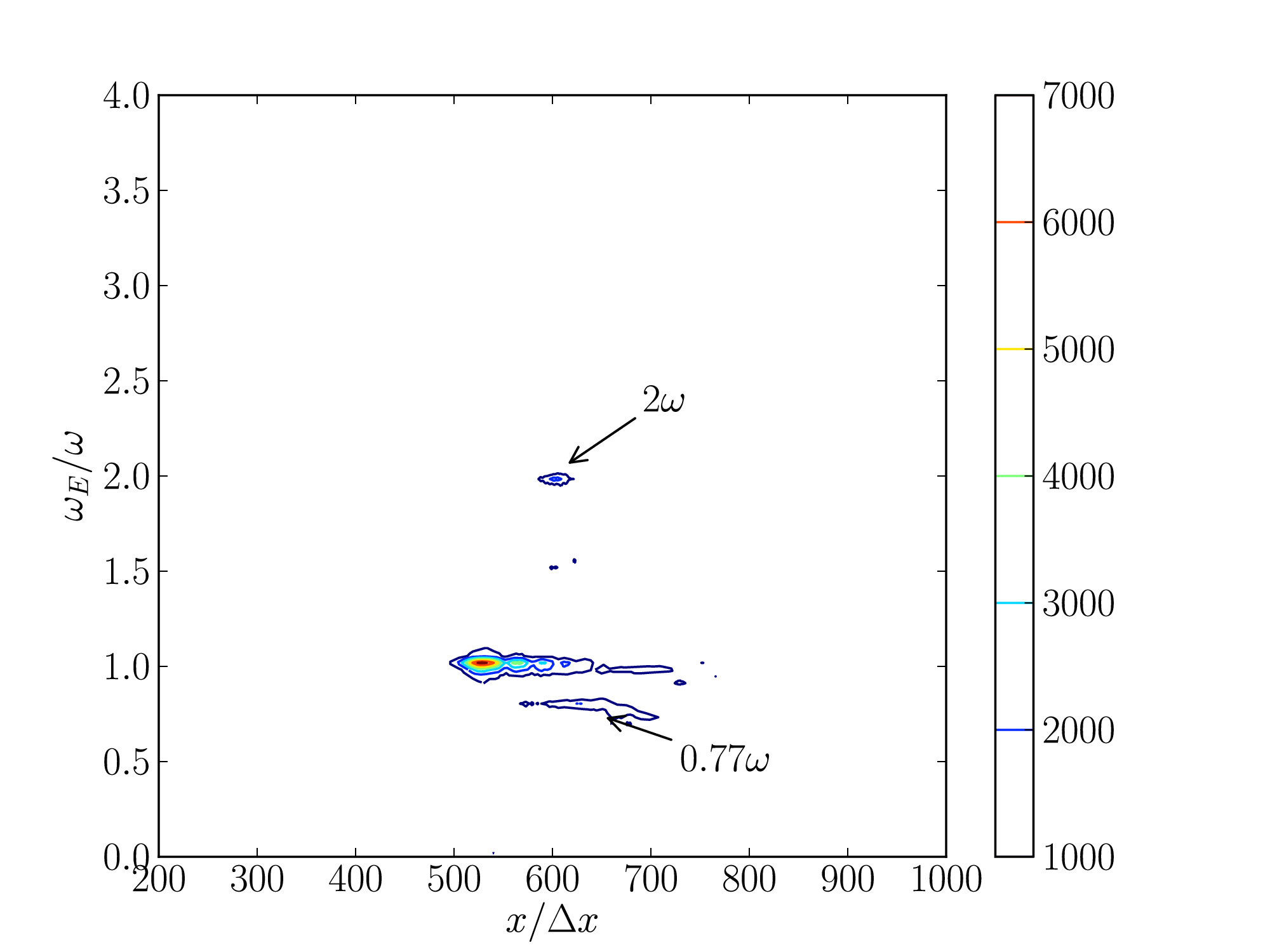}
			\caption{$[406/\omega,580/\omega]$}
			\label{FigSpecEx07}
		\end{subfigure}
		\begin{subfigure}[b]{0.42\textwidth}
			\includegraphics[width=\textwidth]{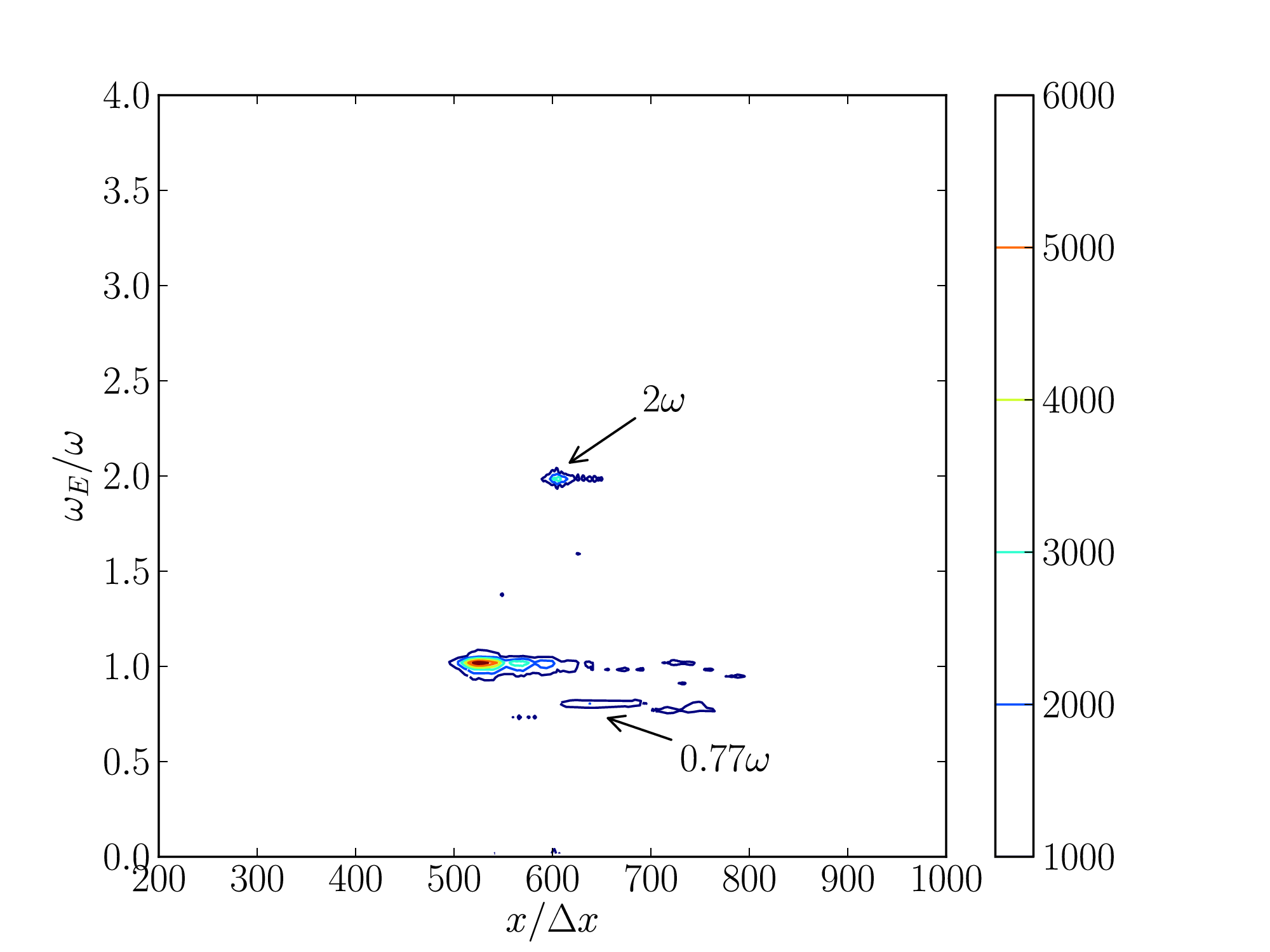}
			\caption{$[580/\omega,754/\omega]$}
			\label{FigSpecEx10}
		\end{subfigure}
		\begin{subfigure}[b]{0.42\textwidth}
			\includegraphics[width=\textwidth]{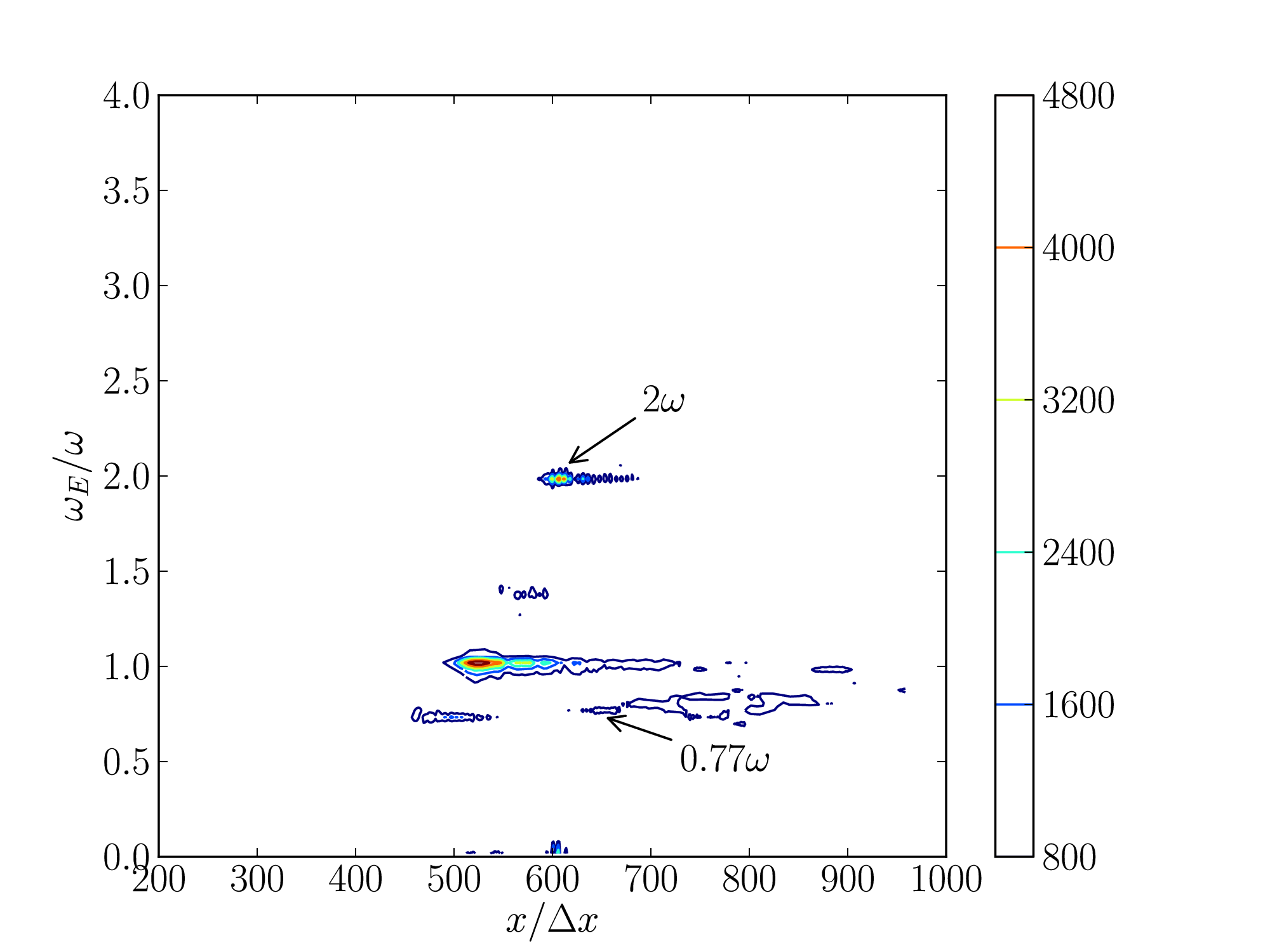}
			\caption{$[812/\omega,986/\omega]$}
			\label{FigSpecEx14}
        	\end{subfigure}
	\end{center}
	\caption{The evolution of frequency spectra of $E_x(x,t)$ during the nonlinear X-B mode conversion. In the simulation $E_1=1.0$~MV/m, $n_r=380$, and the spatial domain of the simulation lies in the interval $[240\Delta x,960\Delta x]$. Each sub-figure denotes the spectrum at different time intervals. We can see that after about $116/\omega$ the wave with the  frequency $2\omega$ is excited, and after $290/\omega$ another wave with frequency $0.77\omega$ appears.}
	\label{FigSpecEx}
\end{figure}

The primary signature of nonlinear processes is
the generation of additional modes with frequencies other
than the incident frequency. In this subsection, 
the new mode generation
during the X-B mode conversion is investigated.
The total number of simulation time steps is set to be 84000.
During the mode conversion, the evolution of spectrum of $E_x$
is recorded in \FIG{FigSpecEx}.

In the initial stage of wave injection,
see \FIG{FigSpecEx00}, there only exists
the wave with frequency $\omega$.
After about $116/\omega$, see \FIG{FigSpecEx02},
the electrostatic wave propagates to the position at $x\sim608\Delta x$,
where a mode with the frequency $2\omega$ appears. 
In Figs.~\ref{FigSpecEx05} and \ref{FigSpecEx07},
the component of the second harmonic mode becomes stronger, 
meanwhile another mode with the frequency $\omega'=0.77\omega$ appears. 
Comparing Figs.~\ref{FigSpecEx10} and \ref{FigSpecEx14}, we 
can see that the mode with frequency $\omega'$ splits into two 
packets, propagating inwards and outwards seperately.

The generation of the mode at frequency 
$2\omega$ can be explained as the self-interaction 
of the initial Bernstein wave.
This second harmonic generation (SHG) of Bernstein wave
has been studied extensively \cite{porkolab2003instabilities,sugawa1988observation,sugaya1989nonlinear,PhysRevLett.100.085002}. Then 
the wave with frequency $\omega'=0.77\omega$ is generated
near the location where the SHG of Bernstein wave 
happens, which implies that the generation of the wave with frequency
$\omega'$ is closely related to the SHG of Bernstein wave.
According to the theory of parametric decay instability (PDI), there 
should be a nother wave with frequency $\omega''$, which satisfies 
either $2\omega + \omega'=\omega''$ or $2\omega - \omega' = \omega''$.
To verify this PDI process,
we plot in \FIG{FigPDI} the dispersion relation of electromagnetic EBWs in hot plasma,
where the electron density of the plasma is
$n_e\left( 608\Delta x \right)$, and the electron temperature is 57.6~eV.

\newcommand{\bfk}{k}
\begin{figure}
	\cpt
	\begin{center}
		\includegraphics[width=0.55\linewidth]{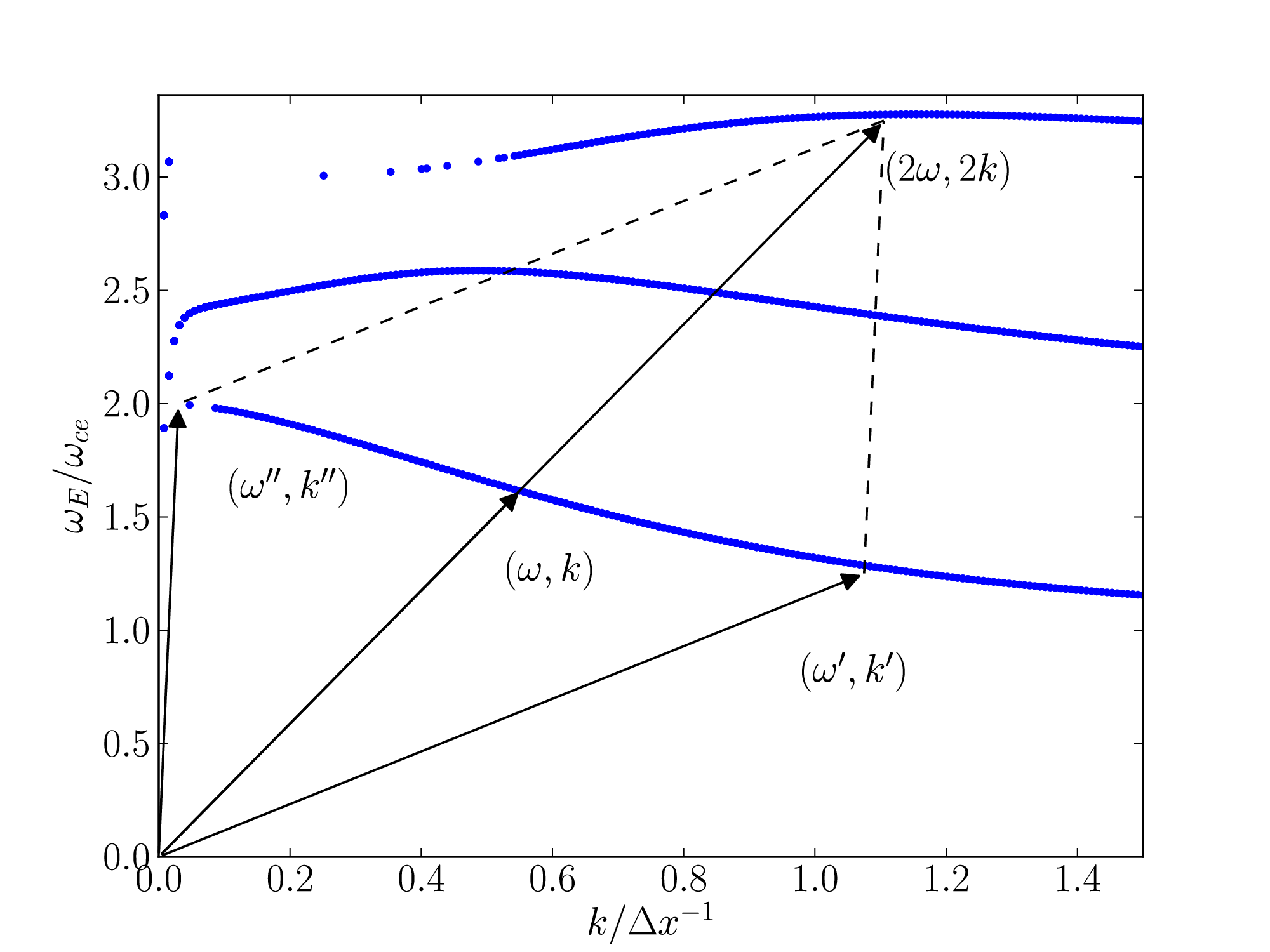}
	\end{center}
	\caption{The dispersion relation of electromagnetic EBWs (blue dots) and wave vectors of the coupled modes (solid arrows). The plasma parameters are the same as those we used in the X-B mode conversion simulation at $x=608\Delta x$ where the wave with frequency $\omega'$ occurs. We can see that these modes satisfy the relation $\bfk''+\bfk'=2\bfk$ and $\omega''+\omega'=2\omega$.}
	\label{FigPDI}
\end{figure}
In \FIG{FigPDI}, it is evident that the frequency and wave number of the third mode satisfy
$\omega''=2\omega-\omega'$ and $k''=2k-k'$, respectively.
This mode can be visualized through the 
spectra of $E_y$ plotted in \FIG{FigSpecEy},
where the mode at frequency $\omega''=1.23\omega$ manifest itself clearly.

\begin{figure}
	\begin{center}
		\begin{subfigure}[b]{0.46\textwidth}
			\includegraphics[width=\textwidth]{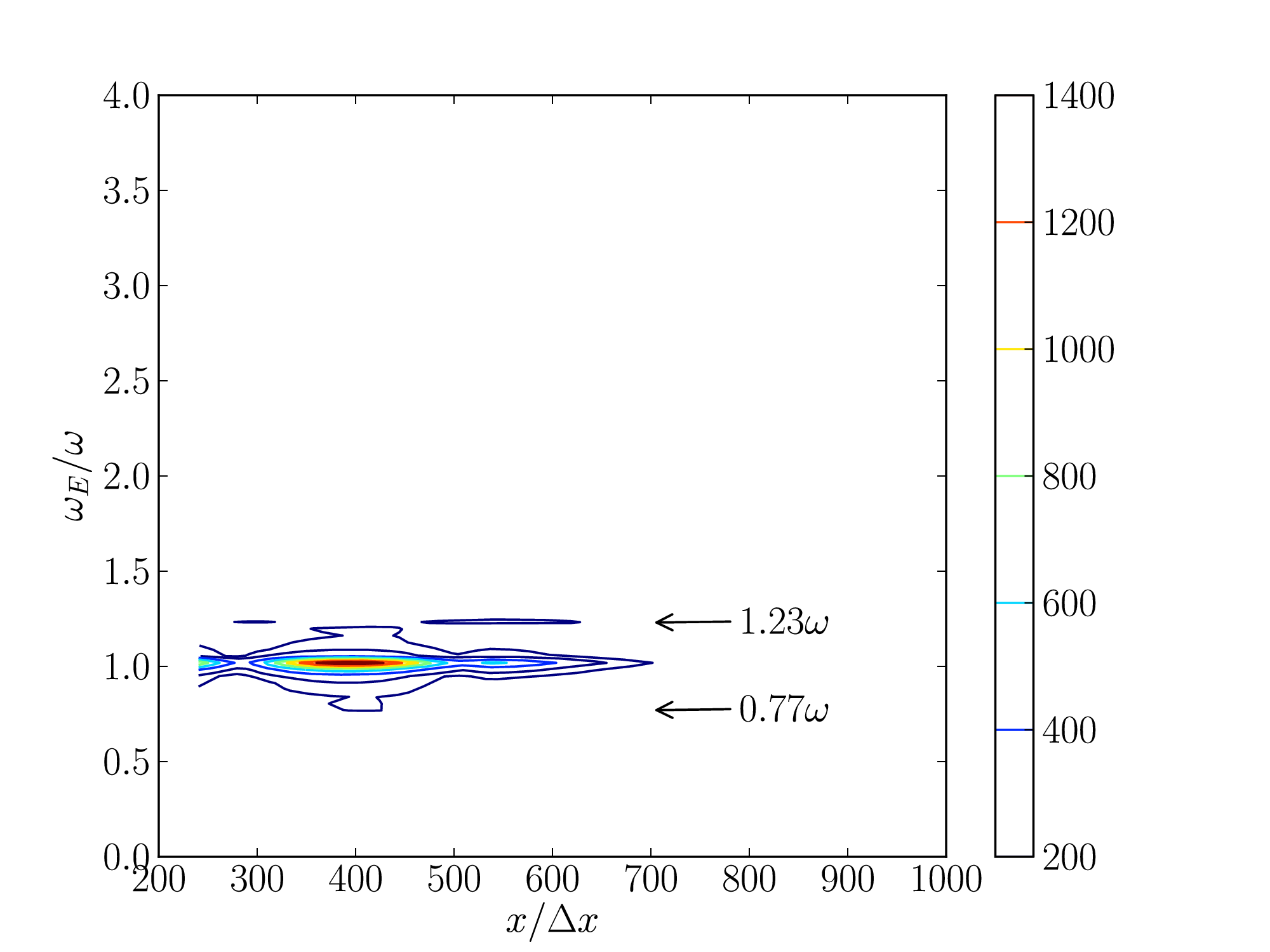}
			\caption{$[580/\omega,754/\omega]$}
			\label{FigSpecEy10}
		\end{subfigure}
		\begin{subfigure}[b]{0.46\textwidth}
			\includegraphics[width=\textwidth]{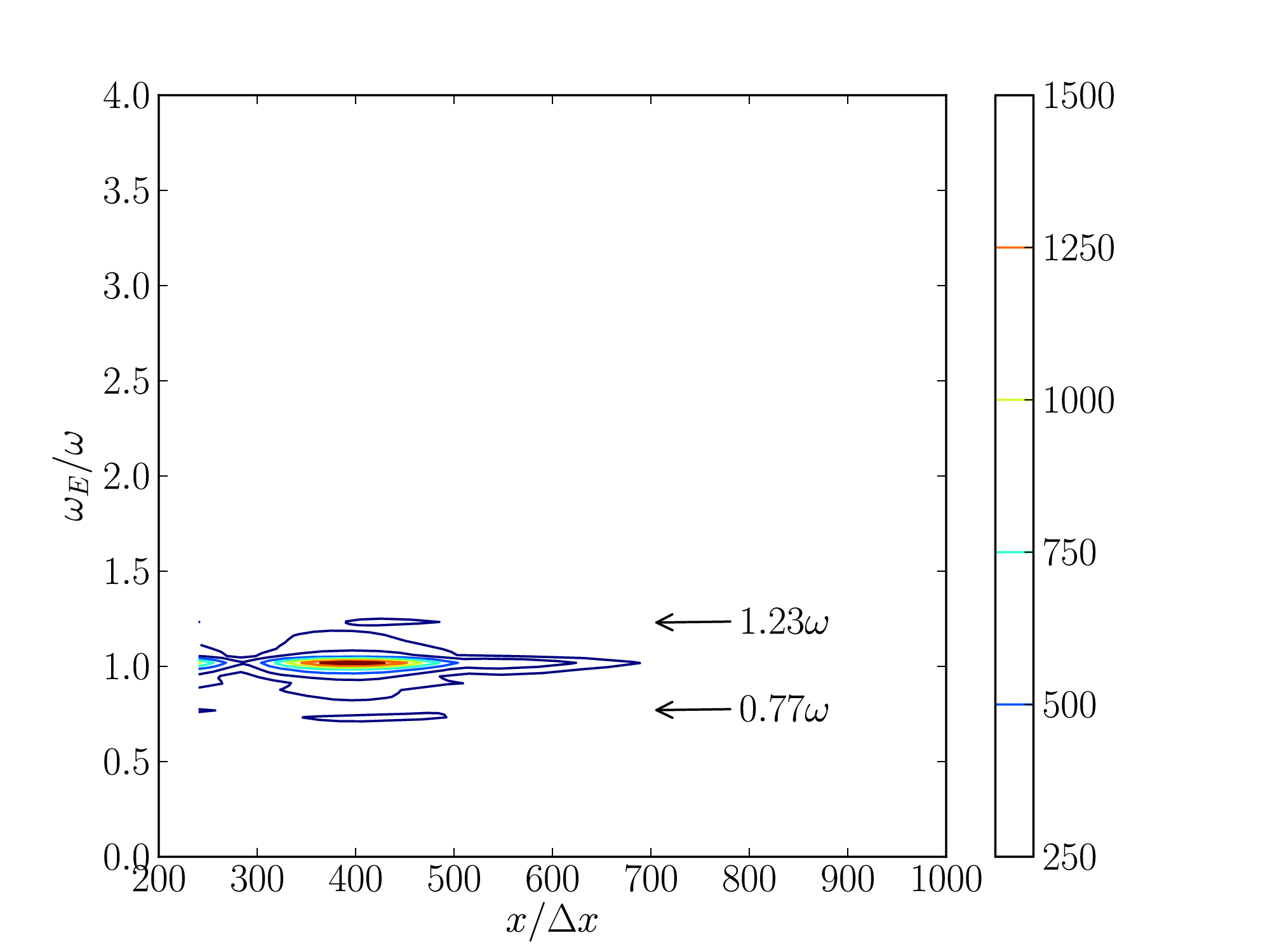}
			\caption{$[812/\omega,986/\omega]$}
			\label{FigSpecEy14}
        	\end{subfigure}
	\end{center}
	\cpt
	\caption{Frequency spectra of $E_y(x,t)$. The scenario is the same as \FIG{FigSpecEx}.}
	\label{FigSpecEy}
\end{figure}
The evolution of $E_x$ spectrum in plasmas with higher electron
temperature (123.9~eV) or weaker incident wave ($E_1 =$0.25~MV/m) are also
investigated. 

When temperature is higher, the location
where the wave with frequency $\omega'=0.77\omega$ appears
shifts to $x\approx680\Delta x$, which is much deeper in the
plasma. This means that the  wave has a 
greater tendency to penetrate the
plasma with higher temperature. At the same time, the reflectivity
is lower.

If we use a smaller source amplitude $E_1$,
the waves at frequencies
$\omega'$ and $\omega''$ generated during the mode conversion
are much weaker than the wave with frequency $\omega$. This is 
of course due to the nonlinear nature of the PDI.

\subsection{Dependence of reflectivity on wave amplitude} \label{SecReflectivities}
In wave heating and current drive the energy conversion efficiency $C$
is a key index. Given the
energy reflectivity $R$,
the conversion efficiency can be calculated according to $C=1-R$.

During the nonlinear X-B mode conversion, the modes with
different frequencies coexist. The reflectivity defined for 
linear case, as in \REF{xiang2006low}, is no
longer valid. In this work, we calculate $R$ based on the Poynting
vectors (see Appendix \ref{SecAPV} for detail).

\begin{figure}
	\cpt
	\begin{center}
		\includegraphics[width=0.5\linewidth]{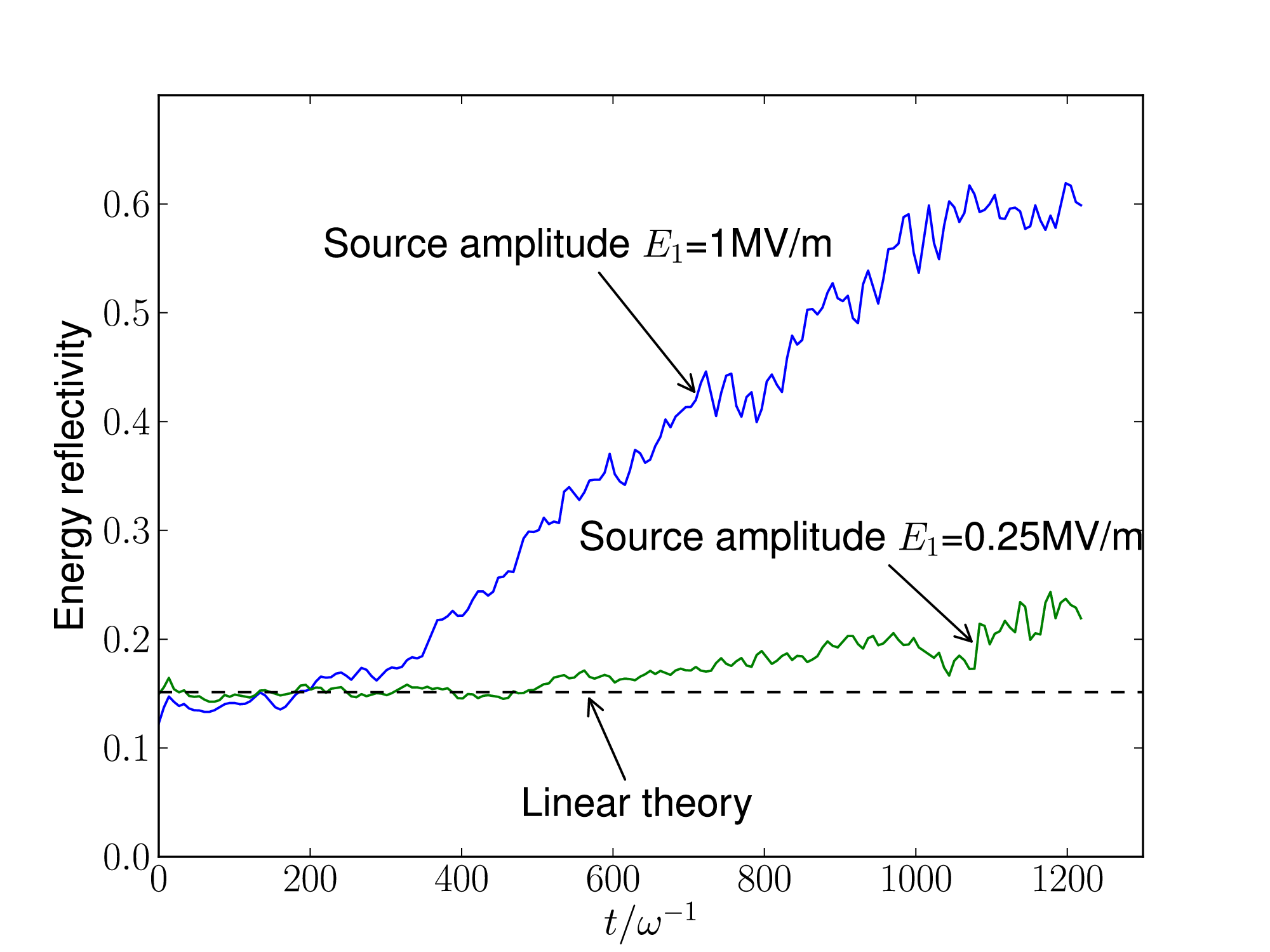}
	\end{center}
	\caption{The evolution of reflectivities when
	waves with different amplitudes were injected into the non-uniform plasma. Here $n_r=380$ and $T_e=57.6$~eV.}
	\label{FigDen95Temp15}
\end{figure}

We now consider the reflectivity of two incident electromagnetic waves with 
different amplitudes, i.e., $E_1=1$~MV/m and $E_1=0.25$~MV/m.
The total number of time steps is 84000, the evolution of 
reflectivities with time are recorded in \FIG{FigDen95Temp15}.
The reflectivity of the wave with the smaller amplitude 
($E_1=0.25$~MV/m) agrees well with the linear mode conversion theory
at the beginning, 
but increases by 10\% after sufficiently long time ($t=1200/\omega$).
For stronger incident wave with amplitude $E_1=1$~MV/m, the
reflectivity increases much faster. 
The reflectivity increases to almost 4 times of its original
value after $t=1200/\omega$.

The simulation results show that the reflectivity is larger 
when the wave amplitude is higher. 
So in realistic wave heating scenario, incident waves with lower power density 
will have larger heating efficiencies. 

\subsection{The energy deposition profile}
\begin{figure}
	\cpt
	\begin{center}
		\begin{subfigure}[b]{0.49\textwidth}
                \includegraphics[width=\textwidth]{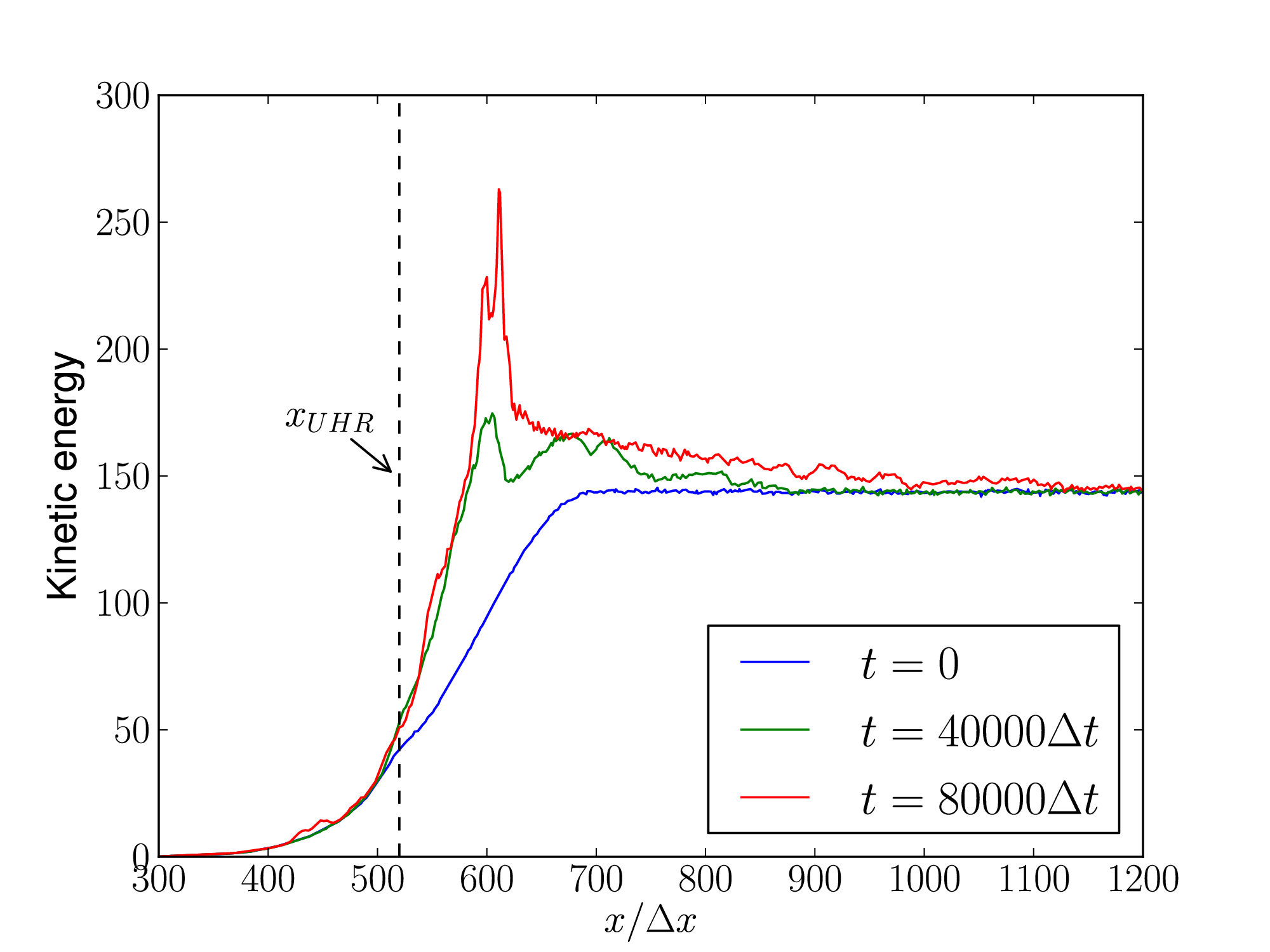}
                	\caption{}
                	\label{FigKinEne6}
        	\end{subfigure}
		\begin{subfigure}[b]{0.49\textwidth}
                \includegraphics[width=\textwidth]{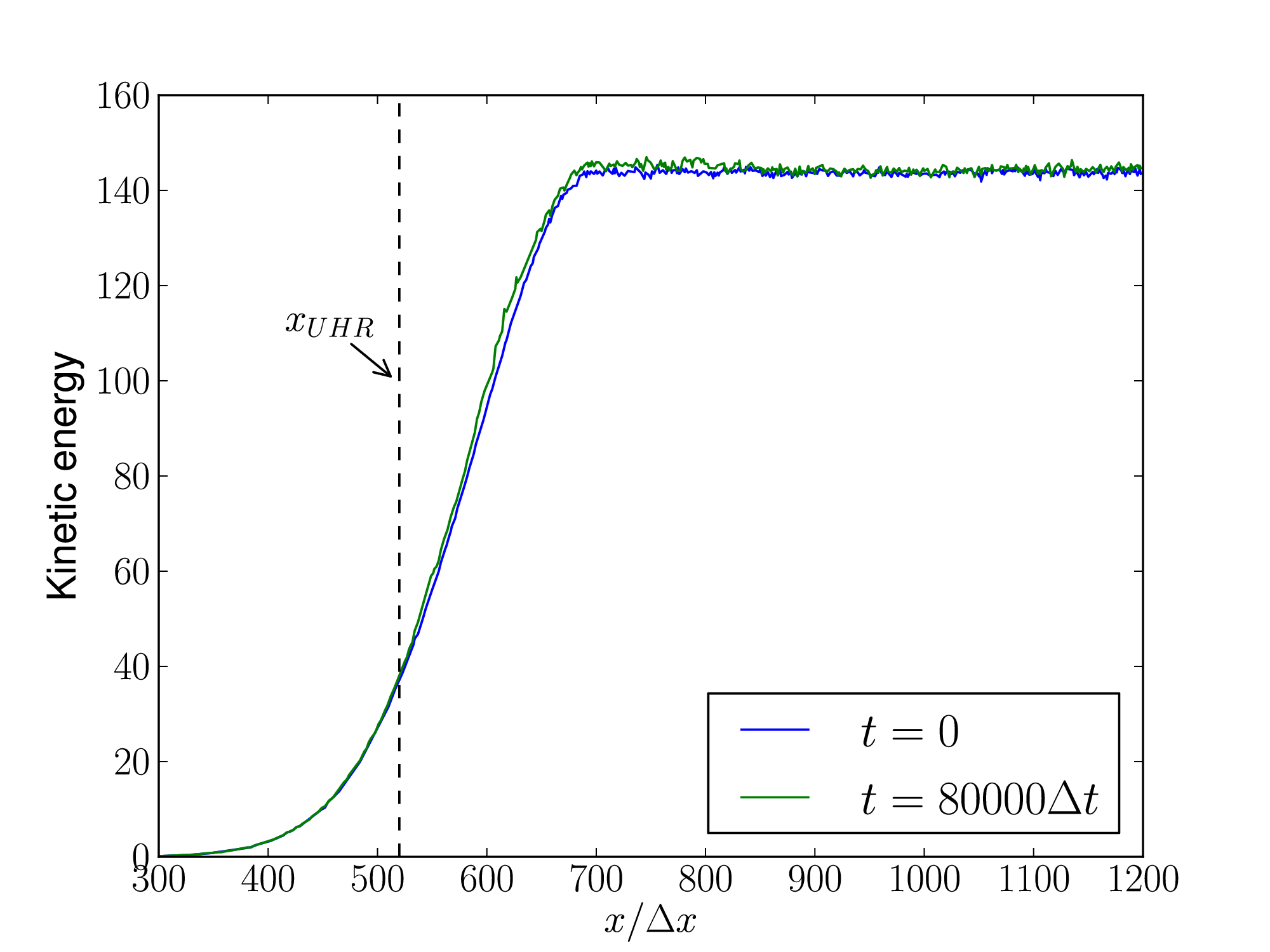}
                	\caption{}
                	\label{FigKinEne2}
        	\end{subfigure}
	\end{center}
	\caption{The difference of electron kinetic energy distribution at the beginning, middle, and end of the simulation. The amplitude of wave sources are 1.0~MV/m in subfigure (a) and 0.25~MV/m in subfigure (b). To suppress the energy oscillation caused by the wave propagation, the kinetic energy are averaged over 4000 time steps.}
	\label{FigKinEne}
\end{figure}

Since the location of wave energy deposition is crucial to rf heating,
it is important to study
the energy deposition profile. Generally speaking, the 
energy depositing inside the plasma core region will be helpful
for heating and current driving. 
In this subsection, the energy
deposition is obtained 
by comparing the kinetic energy distribution of particles
at different times.
In plasmas, the wave energy is stored in both electromagnetic
fields and oscillating particles during each period of the wave.
To calculate the energy deposition rate, we average the energies of particles
over 4000 time steps and record the energy evolution in \FIG{FigKinEne}.

According to \FIG{FigKinEne6}, the energy deposition
happens in a broad range in the plasma if the incident wave
has a larger amplitude.
Most wave energy is deposited near the location 
where the SHG of EBW appears ($x\sim608\Delta x$).
Different from the prediction of the linear theory, both
Figs.~\ref{FigKinEne6} and \ref{FigKinEne2} indicate that the
wave energy penetrates to a deeper position instead of 
accumulating at $x=x_{UHR}$ exactly.
These results can be used for the purpose of increasing the efficiency of auxiliary heating.

The result in \FIG{FigKinEne2}, which shows no evident electron heating, is consistent with the prediction of the linear theory.
According to the linear theory, in the cold plasma approximation the wave energy should accumulate at $x=x_{UHR}$ exactly. 
But in hot plasma, it is expected that the wave energy will be converted into the Bernstein mode and be carried into the plasma interior. 
For perpendicular propagation waves, the only linear damping of the Bernstein wave is caused by relativistic effects, which is weak for 
the electron temperatures in the simulation. So it is difficult for the wave to heat electrons.

On the other hand, for waves with large amplitude in \FIG{FigKinEne6}, 
obvious energy deposition occurs to the right of the UHR position, which is caused by the nonlinear effect. 
Simulation results show that a large amount of wave energy is accumulated at $x\sim608\Delta x$,
where the SHG of EBW appears. From \FIG{FigSpecEx05} and \FIG{FigSpecEy10} it can be observed that the waves with frequencies $\omega'=0.77\omega$ 
and $\omega''=1.23\omega$ are generated at $x\sim608\Delta x$. Notice that $\omega''=1.23\omega\sim2\omega_{ce}$ (see \FIG{FigPDI}), 
and the wave with frequency $\omega''$ will lead to electron heating through the second-harmonic
resonance. The complete process of the nonlinear energy accumulation for large amplitude incident waves is provided as follows.
\begin{enumerate}
	\item The X wave converts to Bernstein wave with the same frequency $\omega$.
	\item The Bernstein wave generates another Bernstein wave with frequency $2\omega$ at $x\sim608\Delta x$ through nonlinear self-interaction.
	\item The new Bernstein wave with $2\omega$ generates two other waves with frequencies $\omega'=0.77\omega$ and $\omega''=1.23\omega\sim2\omega_{ce}$ through PDI.
	\item The wave with $\omega''=1.23\omega\sim2\omega_{ce}$ heats the electrons due to the second-harmonic resonance.
\end{enumerate}
Because the nonlinear effect strongly depends on the parameters of the plasma 
and incident wave, different nonlinear deposition processes will appear
in different situations. We will investigate the large variety of nonlinear
energy deposition phenomena in future studies.

\subsection{Reflectivity influenced by electron temperature}\label{SecIIIA}
\begin{figure}
	\cpt
	\begin{center}
		\includegraphics[width=0.5\linewidth]{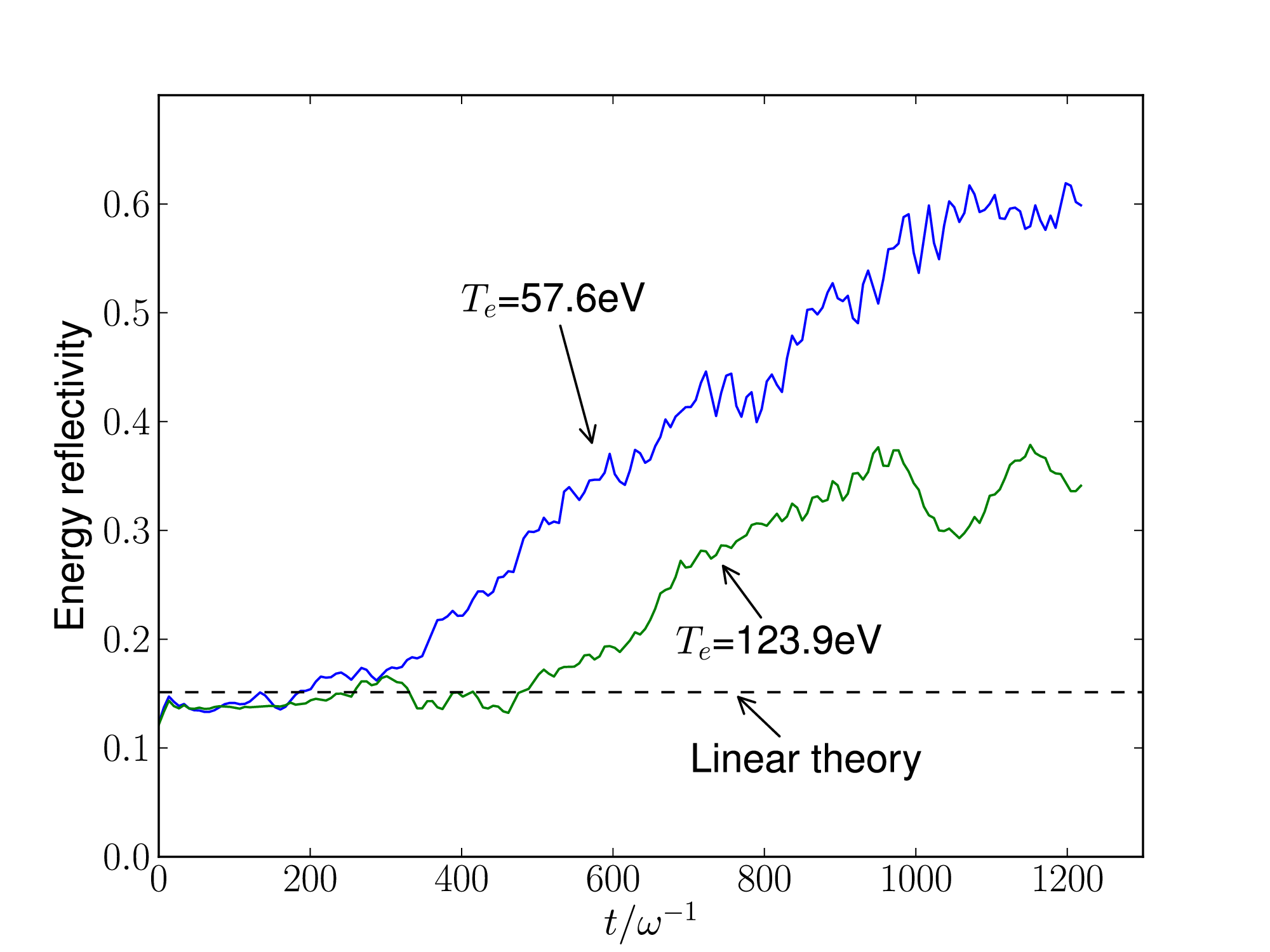}
	\end{center}
	\caption{The reflectivity of waves propagating in plasmas with different
	electron temperatures. Source amplitude $E_1$ is 1.0~MV/m and $n_r=380$.}
	\label{FigCompDen95Temp1522Amp6}
\end{figure}

To find out the dependence of the energy conversion efficiency on the
background
electron temperature,
waves propagating in plasmas with different electron temperatures
are investigated.
The reflectivity evolution in plasmas with electron temperatures $T_e=123.9$~eV and $T_e=57.6$~eV is plotted in \FIG{FigCompDen95Temp1522Amp6}.
We can observe that a long time ($1200/\omega$) after the wave
is injected, the energy reflectivity in the lower $T_e$ is about
twice as large as the higher $T_e$.
This suggests that in wave heating experiments 
a higher electron temperature
should be kept in order to maintain a larger mode conversion efficiency.

\begin{figure}
	\cpt
	\begin{center}
		\includegraphics[width=0.5\linewidth]{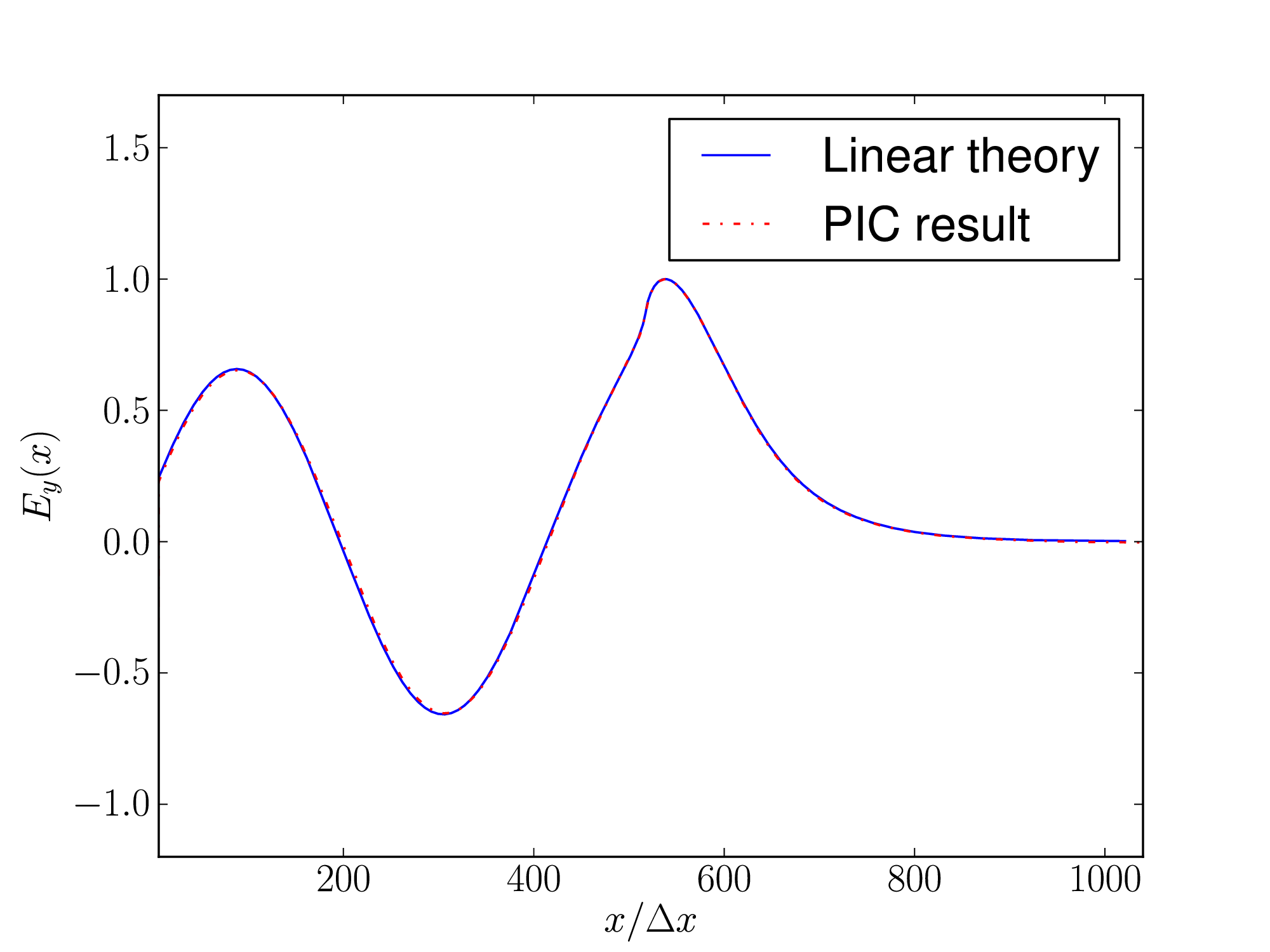}
	\end{center}
	\caption{The spatial distribution $E_y(x)$ of wave injected from the source with small amplitude $E_1=0.066$kV/m at $t=156.484/\omega$ compared with linear theory. Here $n_r=380$ and $T_e\approx0$.}
	\label{FigCompLinPIC}
\end{figure}
\begin{figure}\cpt
	\begin{center}
		\includegraphics[width=0.5\linewidth]{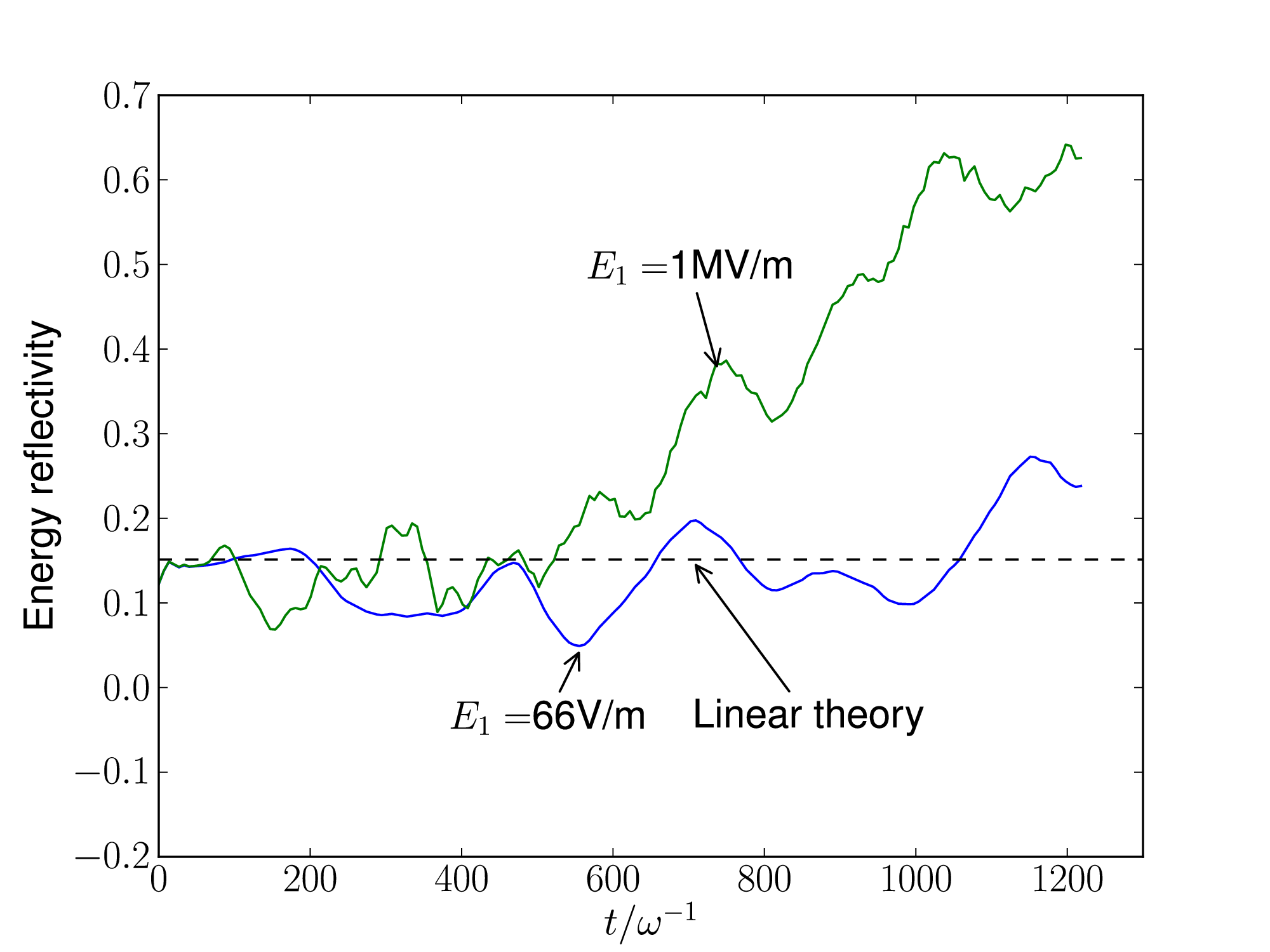}
	\end{center}
	\caption{The energy reflectivity evolution of waves with two different amplitudes for the case of $T_e\approx0$.}
	\label{FigCompAMP1vs6PIC}
\end{figure}

To obtain more information of the relation between
electron temperature and reflectivity,
the case of $T_e\approx$0~eV is also tested.
Two different wave amplitudes $E_1=0.066$kV/m and $E_1=1.0$~MV/m
are studied. 
\FIG{FigCompLinPIC} shows that a short time ($t=156.484/\omega$)
after the wave is injected, $E_y(x)$ of the
wave with small amplitude agrees well with the linear theory.
On the other hand,
\FIG{FigCompAMP1vs6PIC} implies that even for waves with small amplitude,
the mode conversion structure is not stable and the reflectivity
varies significantly after a long time ($t\approx 400/\omega$).
Figure \ref{FigCompAMP1vs6PIC} also suggests that X-B mode conversion
in cold plasmas
for wave with large amplitude ($E_1=$1.0~MV/m) is unstable, and the reflectivity
increases quickly after $t>400/\omega$.

\subsection{Dependence of reflectivity on electron density gradient}

\begin{figure}
	\cpt
	\begin{center}
		\includegraphics[width=0.5\linewidth]{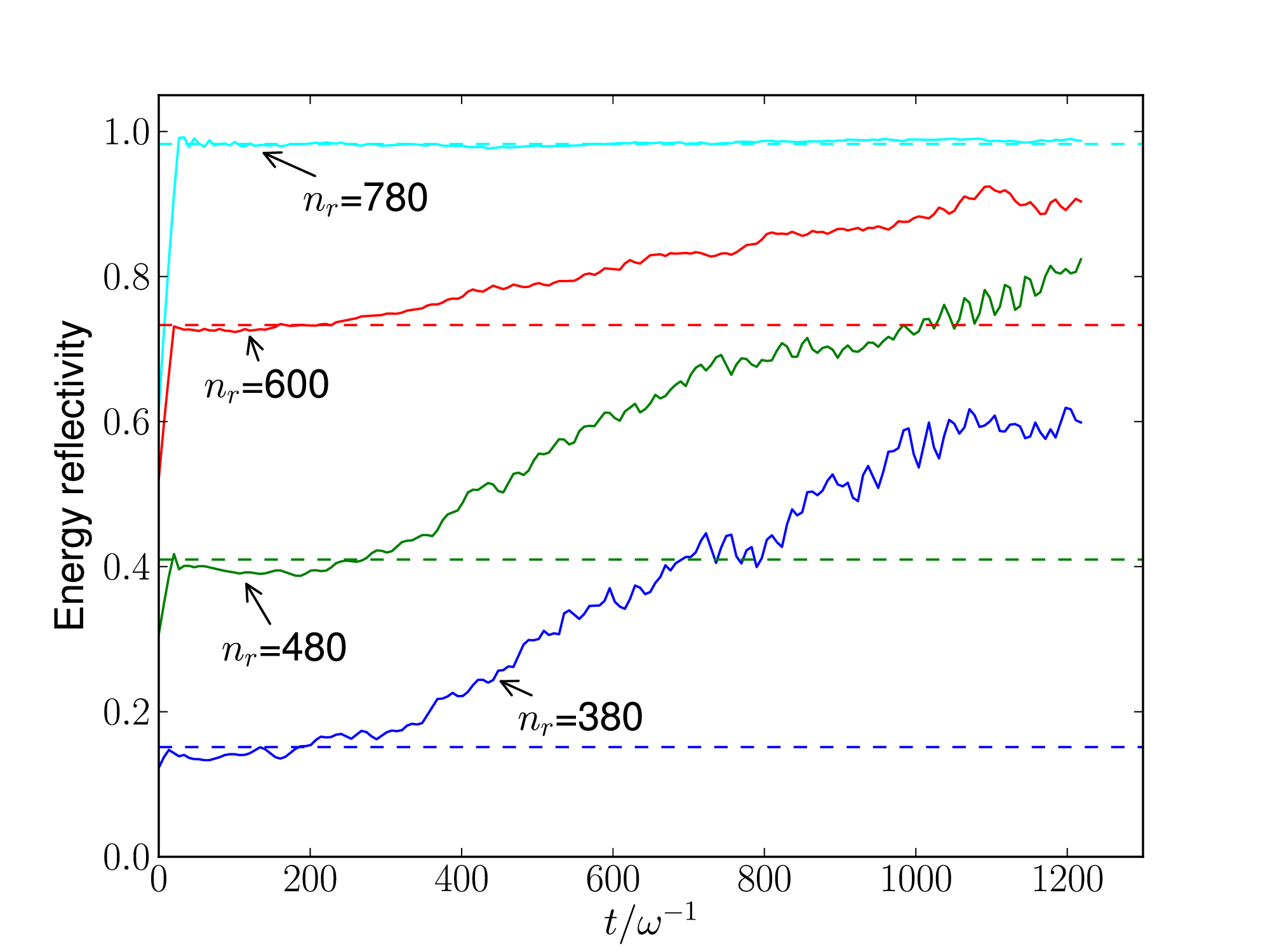}
	\end{center}
	\caption{Comparison of reflectivity with different density
	gradients. Dashed lines are the reflectivity obtained from linear theory. Here $T_e=57.6$~eV.}
	\label{FigRefRateVSDenGrad}
\end{figure}

In this subsection, we investigate the dependence of reflectivity
on the gradient of electron density. Thus we
let $n_r$ in \EQ{EqnDenExample1}
to be \{380, 480, 600, 780\}, $T_e=57.6$~eV, and $E_1=1.0$~MV/m.
The reflectivities with different parameters are
presented in \FIG{FigRefRateVSDenGrad}.
We observe that under all the parameters, the reflectivities 
keep increasing during the mode conversion. The variation of reflectivity 
is very sensitive to the value of density gradient.
Therefore, when applying the X-B rf heating in the magnetic 
confinement devices, it is preferable to maintain a large
edge electron density gradient  
in order to maximize the
conversion efficiency.
\section{Summary and Discussion}\label{SecDiscuss}
The variational symplectic PIC scheme has
been applied to simulate the nonlinear X-B mode
conversion. The preservation of symplectic structure guarantees
the accuracy of the long-term simulation results. 
Simulation results show that when the amplitude
of incident wave is sufficiently large,
the reflectivity will increase during the mode conversion, and
modes with frequencies other than $\omega$ are exited.
Even for waves with small amplitudes, the reflectivity
changes significantly after a long time.
In the meantime, we find that the energy deposits 
within a wide spatial range in the plasma.
The simulation study also suggests that a higher electron temperature can improve
the efficiency of energy conversion. Moreover, a sensitive correlation
between density gradient and reflectivity are observed.

\begin{figure}
	\cpt
	\begin{center}
		\includegraphics[width=0.55\linewidth]{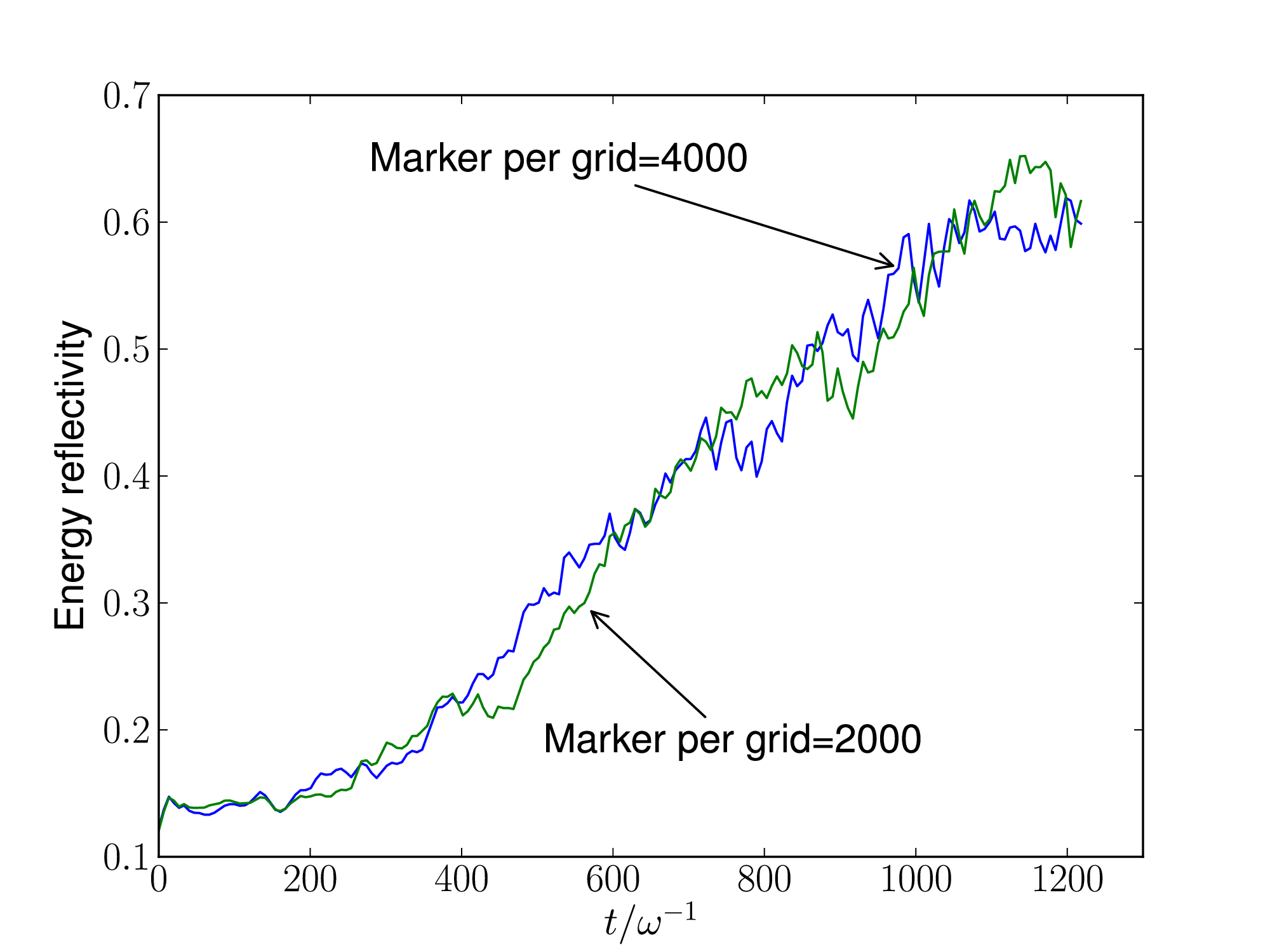}
	\end{center}
	\caption{Comparison of the reflectivity for two cases with different number of sampling particles per grid. It is evident that the physical results are not affected by the number of sampling particles.}
	\label{FigDen95Temp15AMP6PPG}
\end{figure}

In PIC simulations, the numerical noise
always exists due to the relatively less number of 
sampling particles. 
Large numerical noise may lead to inaccurate simulation results.
The accuracy can be verified by checking the influence
of the number of the sampling points on the simulation results.
In the simulation, we have adjusted the 
number of sampling particles per grid from 4000 
to 2000  and repeated the simulation with the same physical parameters to verify that the simulation results are not affected by the numerical noises. See \FIG{FigDen95Temp15AMP6PPG}. 

In this paper, the X-B mode conversion is discussed, and
a series of interesting and useful nonlinear phenomena are 
investigated. The conservation
property of the variational symplectic PIC plays a crucial role 
in achieving the new results during the long-term simulations.
In future studies, 
the long-term simulations of mode conversion of other waves, such as
lower hybrid wave (LHW) and ion cyclotron wave (ICW),
will be carried out. We expect that, the variational symplectic PIC simulation is effective for these
nonlinear rf processes and is able to suggest optimal parameters for rf heating and current drive
schemes.

\appendix
\section{Linear Fluid Theory for the Reflectivity of Extraordinary Wave}\label{SecLinearTheory}

In fluid theory, the evolution of plasma is determined by the following
equations,
\begin{eqnarray}\label{EqnOriginBegin}
	\frac{\partial}{\partial t} n_e &=& -\nabla \cdot \left( n_e\mathbf{u} \right)~,\\
	m_e(\frac{\partial}{\partial t} + \bfu\cdot\nabla)\mathbf{u}&=& e\left( \mathbf{E}+\frac{1}{\mathrm{c}}\mathbf{u}\times \mathbf{B} \right)~,\\
	\frac{1}{\mathrm{c}}\frac{\partial}{\partial t}\bfE&=& \nabla \times\bfB-\frac{4\pi}{\mathrm{c}}\left(e n_e \bfu \right)~,\\\label{EqnOriginEnd}
	\frac{1}{\mathrm{c}}\frac{\partial}{\partial t}\bfB&=& -\nabla\times\bfE~.
\end{eqnarray}

For the propagation of X-waves, we only need to
consider the extraordinary component
($\bfE_1\cdot \mathbf{B}_0 =0$). For 1D inhomogeneous,
field variable can be expressed as
\begin{eqnarray}
	f(\mathbf{r},t)=f_0(x)+f_1(x)\rme^{-\rmi \omega t}~.
\end{eqnarray}
After linearizing Eqs.~(\ref{EqnOriginBegin})-(\ref{EqnOriginEnd}), we obtain a second-order ODE
\begin{eqnarray}\label{EqnFluidODE}
	E_y''(\xi)+Q\left( \xi \right)E_y\left( \xi \right)=0~,
\end{eqnarray}
where
\begin{eqnarray}
	Q\left( \xi \right)&=& 1-\frac{\ope^2\left( \omega^2-\ope^2 \right)}{\omega^2\left( \omega^2-\oce^2-\ope^2 \right)}~,\\
	\xi&=& \frac{x}{\mathrm{c}\omega}~,\\
	\omega_{pe}^2&=& \frac{4\pi n_e(x)e^2}{m_e}~.
\end{eqnarray}
Here $n_e\left( x \right)$ is decided by \EQ{EqnDenExample1}, $n_r=380$, and
the external magnetic field is $\bfB_0=0.55$~T$\bfe_z $.
Note that this ODE may have a singular point $\xi_{UHR}$ where
\begin{eqnarray}
	\ope^2+\oce^2-\omega^2=0~.
\end{eqnarray}
According to previous results of Refs. \cite{budden1988propagation} and \cite{ram1996mode},
the connection condition at $\xi=\xi_{UHR}$ is
\begin{eqnarray}
	E_y|_{\xi_{UHR}-0}&=& E_y|_{\xi_{UHR}+0} = E_y\left( \xi_{UHR} \right)~,\\
	E_y'|_{\xi_{UHR}-0}&=& E_y'|_{\xi_{UHR}+0}-\rmi\pi\beta E_y(\xi_{UHR})~,\\
	\beta&=& \lim_{\xi\rightarrow\xi_{UHR}}\left( \xi-\xi_{UHR} \right)Q\left( \xi \right)~.
\end{eqnarray}
\begin{figure}
	\cpt
	\begin{center}
		\includegraphics[width=0.5\linewidth]{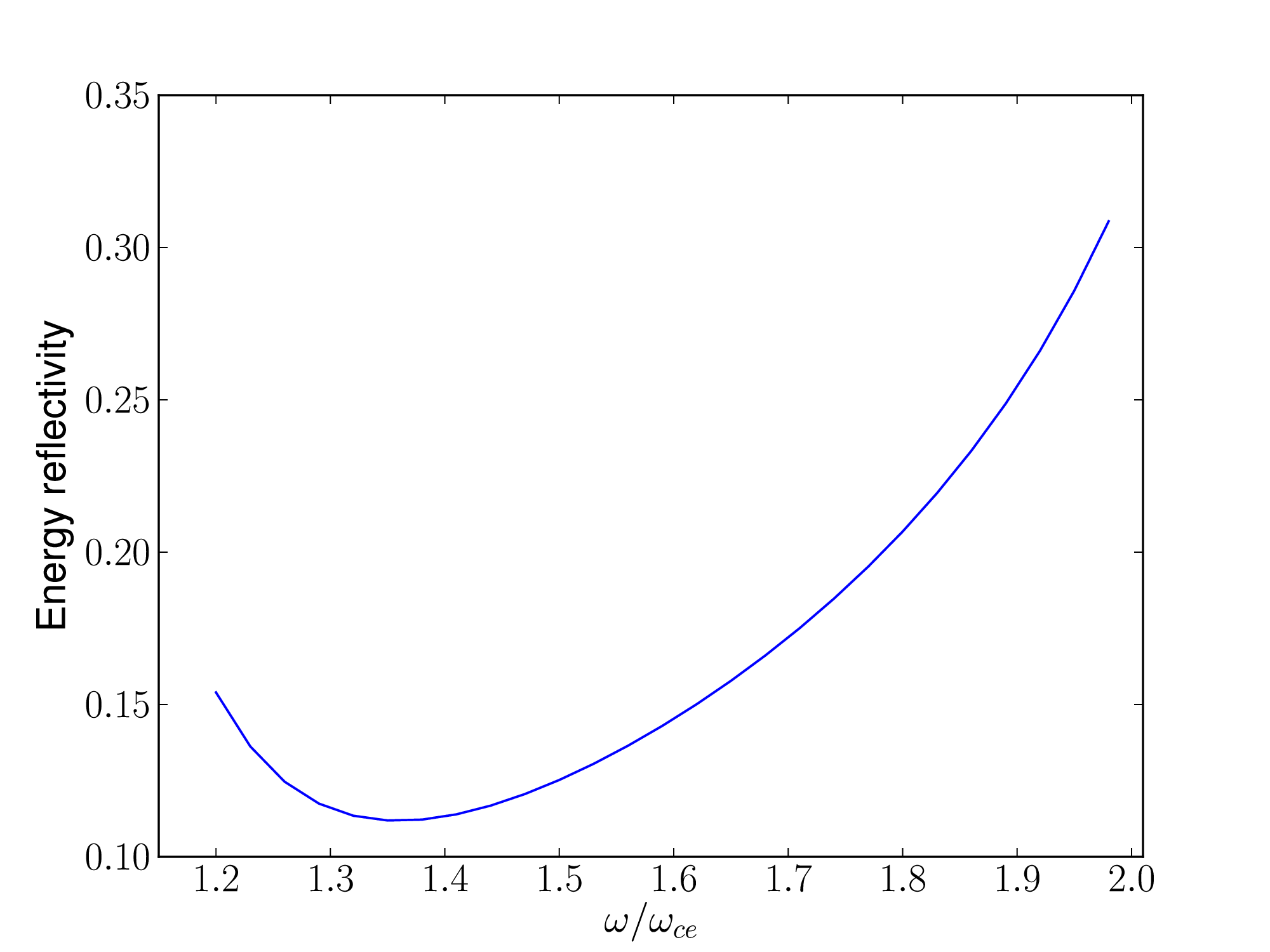}
	\end{center}
	\caption{The relation between amplitude reflectivity and $\omega$, obtained by the linear fluid theory. Here $n_r=380$.}
	\label{FigLinearTheory}
\end{figure}
With these connection conditions, \EQ{EqnFluidODE} can be solved numerically.
To find an appropriate $\omega$ for the incident wave, we calculated the reflectivity for
$\omega \in [1.2\oce,2.0\oce]$ and plotted in \FIG{FigLinearTheory}, which shows
that the lowest reflectivity is at $\omega\approx1.37\oce$. For 
$\omega \in [1.2\oce,2.0\oce]$, the reflectivity will not be too high
to make the mode conversion insignificant. For all our
examples, $\omega$ is set to $1.62\oce$, and the corresponding reflectivity
is $R\approx 0.15$ when $n_r=380$.

\section{The Reflectivity Calculated by Averaged Poynting Vectors}
\label{SecAPV}
The new reflectivity is
defined as follows. Consider the Poynting vector in the front of the
source ($x=8\Delta x$).
The total energy flux $\mathbf{S}_{all}$ consists of incoming and
outcoming parts, which can be expressed as
\begin{eqnarray}
	\mathbf{S}_{all}=\mathbf{S}_{in}+\mathbf{S}_{out}~.
\end{eqnarray}
The incoming part is the energy flux of incoming wave
\begin{eqnarray}
	\mathbf{S}_{in}=\epsilon_0\frac{1}{2}E_1'^2\bfe_z~,
\end{eqnarray}
where $E_1'$ is the amplitude of incoming wave. Note that $E_1'$ is not the amplitude of source $E_1$. In our study, $E_1'=E_1/2$.
The total energy flux is defined as
\begin{eqnarray}
	\mathbf{S}_{all}=\frac{1}{\mu_0}\bfE\times\bfB~,
\end{eqnarray}
and the outcoming part of energy flux is
\begin{eqnarray}
	\mathbf{S}_{out}=\mathbf{S}_{all}-\mathbf{S}_{in}~.
\end{eqnarray}
Therefore, the energy reflectivity is
\begin{eqnarray}
	R&=& \frac{S_{out}}{S_{in}}~.
\end{eqnarray}
All Poynting vectors here are time averaged in
$20\pi/\omega$, which is significantly larger than the period of
waves under investigation.

\begin{acknowledgments}
This research is supported by the National Natural Science Foundation of China (NSFC-11305171),ITER-China Program (2015GB111003,2014GB124005,2013GB111000),JSPS-NRF-NSFC A3 Foresight Program in the field of Plasma Physics (NSFC-11261140328), the CAS Program for Interdisciplinary Collaboration Team, and the GAPS Project.
\end{acknowledgments}

\bibliography{nonlin_XB}

\begin{thebibliography}{26}%
\makeatletter
\providecommand \@ifxundefined [1]{%
 \@ifx{#1\undefined}
}%
\providecommand \@ifnum [1]{%
 \ifnum #1\expandafter \@firstoftwo
 \else \expandafter \@secondoftwo
 \fi
}%
\providecommand \@ifx [1]{%
 \ifx #1\expandafter \@firstoftwo
 \else \expandafter \@secondoftwo
 \fi
}%
\providecommand \natexlab [1]{#1}%
\providecommand \enquote  [1]{``#1''}%
\providecommand \bibnamefont  [1]{#1}%
\providecommand \bibfnamefont [1]{#1}%
\providecommand \citenamefont [1]{#1}%
\providecommand \href@noop [0]{\@secondoftwo}%
\providecommand \href [0]{\begingroup \@sanitize@url \@href}%
\providecommand \@href[1]{\@@startlink{#1}\@@href}%
\providecommand \@@href[1]{\endgroup#1\@@endlink}%
\providecommand \@sanitize@url [0]{\catcode `\\12\catcode `\$12\catcode
  `\&12\catcode `\#12\catcode `\^12\catcode `\_12\catcode `\%12\relax}%
\providecommand \@@startlink[1]{}%
\providecommand \@@endlink[0]{}%
\providecommand \url  [0]{\begingroup\@sanitize@url \@url }%
\providecommand \@url [1]{\endgroup\@href {#1}{\urlprefix }}%
\providecommand \urlprefix  [0]{URL }%
\providecommand \Eprint [0]{\href }%
\providecommand \doibase [0]{http://dx.doi.org/}%
\providecommand \selectlanguage [0]{\@gobble}%
\providecommand \bibinfo  [0]{\@secondoftwo}%
\providecommand \bibfield  [0]{\@secondoftwo}%
\providecommand \translation [1]{[#1]}%
\providecommand \BibitemOpen [0]{}%
\providecommand \bibitemStop [0]{}%
\providecommand \bibitemNoStop [0]{.\EOS\space}%
\providecommand \EOS [0]{\spacefactor3000\relax}%
\providecommand \BibitemShut  [1]{\csname bibitem#1\endcsname}%
\let\auto@bib@innerbib\@empty
\bibitem [{\citenamefont {Ushigusa}\ \emph {et~al.}(1989)\citenamefont
  {Ushigusa}, \citenamefont {Imai}, \citenamefont {Ikeda}, \citenamefont
  {Naito}, \citenamefont {Uehara}, \citenamefont {Yoshida}, \citenamefont
  {Kubo} \emph {et~al.}}]{ushigusa1989lower}%
  \BibitemOpen
  \bibfield  {author} {\bibinfo {author} {\bibfnamefont {K.}~\bibnamefont
  {Ushigusa}}, \bibinfo {author} {\bibfnamefont {T.}~\bibnamefont {Imai}},
  \bibinfo {author} {\bibfnamefont {Y.}~\bibnamefont {Ikeda}}, \bibinfo
  {author} {\bibfnamefont {O.}~\bibnamefont {Naito}}, \bibinfo {author}
  {\bibfnamefont {K.}~\bibnamefont {Uehara}}, \bibinfo {author} {\bibfnamefont
  {H.}~\bibnamefont {Yoshida}}, \bibinfo {author} {\bibfnamefont
  {H.}~\bibnamefont {Kubo}},  \emph {et~al.},\ }\href@noop {} {\bibfield
  {journal} {\bibinfo  {journal} {Nucl. Fusion}\ }\textbf {\bibinfo {volume}
  {29}},\ \bibinfo {pages} {1052} (\bibinfo {year} {1989})}\BibitemShut
  {NoStop}%
\bibitem [{\citenamefont {Lianmin}\ \emph {et~al.}(2010)\citenamefont
  {Lianmin}, \citenamefont {Jiafang}, \citenamefont {Fukun}, \citenamefont
  {Hua}, \citenamefont {Mao}, \citenamefont {Liang}, \citenamefont {Xiaojie},\
  and\ \citenamefont {Handong}}]{lianmin20102450}%
  \BibitemOpen
  \bibfield  {author} {\bibinfo {author} {\bibfnamefont {Z.}~\bibnamefont
  {Lianmin}}, \bibinfo {author} {\bibfnamefont {S.}~\bibnamefont {Jiafang}},
  \bibinfo {author} {\bibfnamefont {L.}~\bibnamefont {Fukun}}, \bibinfo
  {author} {\bibfnamefont {J.}~\bibnamefont {Hua}}, \bibinfo {author}
  {\bibfnamefont {W.}~\bibnamefont {Mao}}, \bibinfo {author} {\bibfnamefont
  {L.}~\bibnamefont {Liang}}, \bibinfo {author} {\bibfnamefont
  {W.}~\bibnamefont {Xiaojie}}, \ and\ \bibinfo {author} {\bibfnamefont
  {X.}~\bibnamefont {Handong}},\ }\href@noop {} {\bibfield  {journal} {\bibinfo
   {journal} {Plasma Sci. Technol.}\ }\textbf {\bibinfo {volume} {12}},\
  \bibinfo {pages} {118} (\bibinfo {year} {2010})}\BibitemShut {NoStop}%
\bibitem [{\citenamefont {Start}\ \emph {et~al.}(1998)\citenamefont {Start},
  \citenamefont {Jacquinot}, \citenamefont {Bergeaud}, \citenamefont
  {Bhatnagar}, \citenamefont {Cottrell}, \citenamefont {Clement}, \citenamefont
  {Eriksson}, \citenamefont {Fasoli}, \citenamefont {Gondhalekar},
  \citenamefont {Gormezano} \emph {et~al.}}]{start1998dt}%
  \BibitemOpen
  \bibfield  {author} {\bibinfo {author} {\bibfnamefont {D.}~\bibnamefont
  {Start}}, \bibinfo {author} {\bibfnamefont {J.}~\bibnamefont {Jacquinot}},
  \bibinfo {author} {\bibfnamefont {V.}~\bibnamefont {Bergeaud}}, \bibinfo
  {author} {\bibfnamefont {V.}~\bibnamefont {Bhatnagar}}, \bibinfo {author}
  {\bibfnamefont {G.}~\bibnamefont {Cottrell}}, \bibinfo {author}
  {\bibfnamefont {S.}~\bibnamefont {Clement}}, \bibinfo {author} {\bibfnamefont
  {L.}~\bibnamefont {Eriksson}}, \bibinfo {author} {\bibfnamefont
  {A.}~\bibnamefont {Fasoli}}, \bibinfo {author} {\bibfnamefont
  {A.}~\bibnamefont {Gondhalekar}}, \bibinfo {author} {\bibfnamefont
  {C.}~\bibnamefont {Gormezano}},  \emph {et~al.},\ }\href@noop {} {\bibfield
  {journal} {\bibinfo  {journal} {Phys. Rev. Lett.}\ }\textbf {\bibinfo
  {volume} {80}},\ \bibinfo {pages} {4681} (\bibinfo {year}
  {1998})}\BibitemShut {NoStop}%
\bibitem [{\citenamefont {Laqua}\ \emph {et~al.}(1997)\citenamefont {Laqua},
  \citenamefont {Erckmann}, \citenamefont {Hartfu{\ss}}, \citenamefont {Laqua}
  \emph {et~al.}}]{laqua1997resonant}%
  \BibitemOpen
  \bibfield  {author} {\bibinfo {author} {\bibfnamefont {H.}~\bibnamefont
  {Laqua}}, \bibinfo {author} {\bibfnamefont {V.}~\bibnamefont {Erckmann}},
  \bibinfo {author} {\bibfnamefont {H.}~\bibnamefont {Hartfu{\ss}}}, \bibinfo
  {author} {\bibfnamefont {H.}~\bibnamefont {Laqua}},  \emph {et~al.},\
  }\href@noop {} {\bibfield  {journal} {\bibinfo  {journal} {Phys. Rev. Lett.}\
  }\textbf {\bibinfo {volume} {78}},\ \bibinfo {pages} {3467} (\bibinfo {year}
  {1997})}\BibitemShut {NoStop}%
\bibitem [{\citenamefont {Cairns}\ and\ \citenamefont
  {Lashmore-Davies}(2000)}]{cairns2000prospects}%
  \BibitemOpen
  \bibfield  {author} {\bibinfo {author} {\bibfnamefont {R.}~\bibnamefont
  {Cairns}}\ and\ \bibinfo {author} {\bibfnamefont {C.}~\bibnamefont
  {Lashmore-Davies}},\ }\href@noop {} {\bibfield  {journal} {\bibinfo
  {journal} {Phys. Plasmas}\ }\textbf {\bibinfo {volume} {7}},\ \bibinfo
  {pages} {4126} (\bibinfo {year} {2000})}\BibitemShut {NoStop}%
\bibitem [{\citenamefont {Kuehl}(1967)}]{kuehl1967coupling}%
  \BibitemOpen
  \bibfield  {author} {\bibinfo {author} {\bibfnamefont {H.}~\bibnamefont
  {Kuehl}},\ }\href@noop {} {\bibfield  {journal} {\bibinfo  {journal} {Phys.
  Rev.}\ }\textbf {\bibinfo {volume} {154}},\ \bibinfo {pages} {124} (\bibinfo
  {year} {1967})}\BibitemShut {NoStop}%
\bibitem [{\citenamefont {Ram}\ and\ \citenamefont
  {Schultz}(2000)}]{ram2000excitation}%
  \BibitemOpen
  \bibfield  {author} {\bibinfo {author} {\bibfnamefont {A.}~\bibnamefont
  {Ram}}\ and\ \bibinfo {author} {\bibfnamefont {S.~D.}\ \bibnamefont
  {Schultz}},\ }\href@noop {} {\bibfield  {journal} {\bibinfo  {journal} {Phys.
  Plasmas}\ }\textbf {\bibinfo {volume} {7}},\ \bibinfo {pages} {4084}
  (\bibinfo {year} {2000})}\BibitemShut {NoStop}%
\bibitem [{\citenamefont {Jones}\ \emph {et~al.}(2003)\citenamefont {Jones},
  \citenamefont {Efthimion}, \citenamefont {Taylor}, \citenamefont {Munsat},
  \citenamefont {Wilson}, \citenamefont {Hosea}, \citenamefont {Kaita},
  \citenamefont {Majeski}, \citenamefont {Maingi}, \citenamefont {Shiraiwa}
  \emph {et~al.}}]{jones2003controlled}%
  \BibitemOpen
  \bibfield  {author} {\bibinfo {author} {\bibfnamefont {B.}~\bibnamefont
  {Jones}}, \bibinfo {author} {\bibfnamefont {P.}~\bibnamefont {Efthimion}},
  \bibinfo {author} {\bibfnamefont {G.}~\bibnamefont {Taylor}}, \bibinfo
  {author} {\bibfnamefont {T.}~\bibnamefont {Munsat}}, \bibinfo {author}
  {\bibfnamefont {J.}~\bibnamefont {Wilson}}, \bibinfo {author} {\bibfnamefont
  {J.}~\bibnamefont {Hosea}}, \bibinfo {author} {\bibfnamefont
  {R.}~\bibnamefont {Kaita}}, \bibinfo {author} {\bibfnamefont
  {R.}~\bibnamefont {Majeski}}, \bibinfo {author} {\bibfnamefont
  {R.}~\bibnamefont {Maingi}}, \bibinfo {author} {\bibfnamefont
  {S.}~\bibnamefont {Shiraiwa}},  \emph {et~al.},\ }\href@noop {} {\bibfield
  {journal} {\bibinfo  {journal} {Phys. Rev. Lett.}\ }\textbf {\bibinfo
  {volume} {90}},\ \bibinfo {pages} {165001} (\bibinfo {year}
  {2003})}\BibitemShut {NoStop}%
\bibitem [{\citenamefont {Shiraiwa}\ \emph {et~al.}(2006)\citenamefont
  {Shiraiwa}, \citenamefont {Hanada}, \citenamefont {Hasegawa}, \citenamefont
  {Idei}, \citenamefont {Kasahara}, \citenamefont {Mitarai}, \citenamefont
  {Nakamura}, \citenamefont {Nishino}, \citenamefont {Nozato}, \citenamefont
  {Sakamoto} \emph {et~al.}}]{shiraiwa2006heating}%
  \BibitemOpen
  \bibfield  {author} {\bibinfo {author} {\bibfnamefont {S.}~\bibnamefont
  {Shiraiwa}}, \bibinfo {author} {\bibfnamefont {K.}~\bibnamefont {Hanada}},
  \bibinfo {author} {\bibfnamefont {M.}~\bibnamefont {Hasegawa}}, \bibinfo
  {author} {\bibfnamefont {H.}~\bibnamefont {Idei}}, \bibinfo {author}
  {\bibfnamefont {H.}~\bibnamefont {Kasahara}}, \bibinfo {author}
  {\bibfnamefont {O.}~\bibnamefont {Mitarai}}, \bibinfo {author} {\bibfnamefont
  {K.}~\bibnamefont {Nakamura}}, \bibinfo {author} {\bibfnamefont
  {N.}~\bibnamefont {Nishino}}, \bibinfo {author} {\bibfnamefont
  {H.}~\bibnamefont {Nozato}}, \bibinfo {author} {\bibfnamefont
  {M.}~\bibnamefont {Sakamoto}},  \emph {et~al.},\ }\href@noop {} {\bibfield
  {journal} {\bibinfo  {journal} {Phys. Rev. Lett.}\ }\textbf {\bibinfo
  {volume} {96}},\ \bibinfo {pages} {185003} (\bibinfo {year}
  {2006})}\BibitemShut {NoStop}%
\bibitem [{\citenamefont {Budden}(1988)}]{budden1988propagation}%
  \BibitemOpen
  \bibfield  {author} {\bibinfo {author} {\bibfnamefont {K.~G.}\ \bibnamefont
  {Budden}},\ }\href@noop {} {\emph {\bibinfo {title} {The propagation of radio
  waves: the theory of radio waves of low power in the ionosphere and
  magnetosphere}}}\ (\bibinfo  {publisher} {Cambridge University Press},\
  \bibinfo {year} {1988})\ pp.\ \bibinfo {pages} {596--602}\BibitemShut
  {NoStop}%
\bibitem [{\citenamefont {Ram}\ \emph {et~al.}(1996)\citenamefont {Ram},
  \citenamefont {Bers}, \citenamefont {Schultz},\ and\ \citenamefont
  {Fuchs}}]{ram1996mode}%
  \BibitemOpen
  \bibfield  {author} {\bibinfo {author} {\bibfnamefont {A.}~\bibnamefont
  {Ram}}, \bibinfo {author} {\bibfnamefont {A.}~\bibnamefont {Bers}}, \bibinfo
  {author} {\bibfnamefont {S.}~\bibnamefont {Schultz}}, \ and\ \bibinfo
  {author} {\bibfnamefont {V.}~\bibnamefont {Fuchs}},\ }\href@noop {}
  {\bibfield  {journal} {\bibinfo  {journal} {Phys. Plasmas}\ }\textbf
  {\bibinfo {volume} {3}},\ \bibinfo {pages} {1976} (\bibinfo {year}
  {1996})}\BibitemShut {NoStop}%
\bibitem [{\citenamefont {Sugawa}(1988)}]{sugawa1988observation}%
  \BibitemOpen
  \bibfield  {author} {\bibinfo {author} {\bibfnamefont {M.}~\bibnamefont
  {Sugawa}},\ }\href@noop {} {\bibfield  {journal} {\bibinfo  {journal} {Phys.
  Rev. Lett.}\ }\textbf {\bibinfo {volume} {61}},\ \bibinfo {pages} {543}
  (\bibinfo {year} {1988})}\BibitemShut {NoStop}%
\bibitem [{\citenamefont {Porkolab}(1985)}]{porkolab1985nonlinear}%
  \BibitemOpen
  \bibfield  {author} {\bibinfo {author} {\bibfnamefont {M.}~\bibnamefont
  {Porkolab}},\ }\href@noop {} {\bibfield  {journal} {\bibinfo  {journal}
  {Phys. Rev. Lett.}\ }\textbf {\bibinfo {volume} {54}},\ \bibinfo {pages}
  {434} (\bibinfo {year} {1985})}\BibitemShut {NoStop}%
\bibitem [{\citenamefont {Xiang}\ \emph {et~al.}(2006)\citenamefont {Xiang},
  \citenamefont {Cary}, \citenamefont {Barnes},\ and\ \citenamefont
  {Carlsson}}]{xiang2006low}%
  \BibitemOpen
  \bibfield  {author} {\bibinfo {author} {\bibfnamefont {N.}~\bibnamefont
  {Xiang}}, \bibinfo {author} {\bibfnamefont {J.~R.}\ \bibnamefont {Cary}},
  \bibinfo {author} {\bibfnamefont {D.~C.}\ \bibnamefont {Barnes}}, \ and\
  \bibinfo {author} {\bibfnamefont {J.}~\bibnamefont {Carlsson}},\ }\href@noop
  {} {\bibfield  {journal} {\bibinfo  {journal} {Phys. Plasmas}\ }\textbf
  {\bibinfo {volume} {13}},\ \bibinfo {pages} {062111} (\bibinfo {year}
  {2006})}\BibitemShut {NoStop}%
\bibitem [{\citenamefont {Yu}\ and\ \citenamefont
  {Qin}(2009)}]{yu2009gyrocenter}%
  \BibitemOpen
  \bibfield  {author} {\bibinfo {author} {\bibfnamefont {Z.}~\bibnamefont
  {Yu}}\ and\ \bibinfo {author} {\bibfnamefont {H.}~\bibnamefont {Qin}},\
  }\href@noop {} {\bibfield  {journal} {\bibinfo  {journal} {Phys. Plasmas}\
  }\textbf {\bibinfo {volume} {16}},\ \bibinfo {pages} {032507} (\bibinfo
  {year} {2009})}\BibitemShut {NoStop}%
\bibitem [{\citenamefont {Liu}\ \emph {et~al.}(2014)\citenamefont {Liu},
  \citenamefont {Yu},\ and\ \citenamefont {Qin}}]{liunonlinear}%
  \BibitemOpen
  \bibfield  {author} {\bibinfo {author} {\bibfnamefont {J.}~\bibnamefont
  {Liu}}, \bibinfo {author} {\bibfnamefont {Z.}~\bibnamefont {Yu}}, \ and\
  \bibinfo {author} {\bibfnamefont {H.}~\bibnamefont {Qin}},\ }\href@noop {}
  {\bibfield  {journal} {\bibinfo  {journal} {Commun. Comput. Phys.}\ }\textbf
  {\bibinfo {volume} {15}},\ \bibinfo {pages} {1167} (\bibinfo {year}
  {2014})}\BibitemShut {NoStop}%
\bibitem [{\citenamefont {Xiang}\ and\ \citenamefont
  {Cary}(2008)}]{PhysRevLett.100.085002}%
  \BibitemOpen
  \bibfield  {author} {\bibinfo {author} {\bibfnamefont {N.}~\bibnamefont
  {Xiang}}\ and\ \bibinfo {author} {\bibfnamefont {J.~R.}\ \bibnamefont
  {Cary}},\ }\href {\doibase 10.1103/PhysRevLett.100.085002} {\bibfield
  {journal} {\bibinfo  {journal} {Phys. Rev. Lett.}\ }\textbf {\bibinfo
  {volume} {100}},\ \bibinfo {pages} {085002} (\bibinfo {year}
  {2008})}\BibitemShut {NoStop}%
\bibitem [{\citenamefont {Xiao}\ \emph {et~al.}(2013)\citenamefont {Xiao},
  \citenamefont {Liu}, \citenamefont {Qin},\ and\ \citenamefont
  {Yu}}]{xiao2013variational}%
  \BibitemOpen
  \bibfield  {author} {\bibinfo {author} {\bibfnamefont {J.}~\bibnamefont
  {Xiao}}, \bibinfo {author} {\bibfnamefont {J.}~\bibnamefont {Liu}}, \bibinfo
  {author} {\bibfnamefont {H.}~\bibnamefont {Qin}}, \ and\ \bibinfo {author}
  {\bibfnamefont {Z.}~\bibnamefont {Yu}},\ }\href@noop {} {\bibfield  {journal}
  {\bibinfo  {journal} {Phys. Plasmas}\ }\textbf {\bibinfo {volume} {20}},\
  \bibinfo {pages} {102517} (\bibinfo {year} {2013})}\BibitemShut {NoStop}%
\bibitem [{\citenamefont {Hairer}\ \emph {et~al.}(2006)\citenamefont {Hairer},
  \citenamefont {Lubich},\ and\ \citenamefont {Wanner}}]{hairer2006geometric}%
  \BibitemOpen
  \bibfield  {author} {\bibinfo {author} {\bibfnamefont {E.}~\bibnamefont
  {Hairer}}, \bibinfo {author} {\bibfnamefont {C.}~\bibnamefont {Lubich}}, \
  and\ \bibinfo {author} {\bibfnamefont {G.}~\bibnamefont {Wanner}},\
  }\href@noop {} {\emph {\bibinfo {title} {Geometric numerical integration:
  structure-preserving algorithms for ordinary differential equations}}},\
  Vol.~\bibinfo {volume} {31}\ (\bibinfo  {publisher} {Springer},\ \bibinfo
  {year} {2006})\ pp.\ \bibinfo {pages} {389--434}\BibitemShut {NoStop}%
\bibitem [{\citenamefont {Marsden}\ and\ \citenamefont
  {West}(2001)}]{marsden2001discrete}%
  \BibitemOpen
  \bibfield  {author} {\bibinfo {author} {\bibfnamefont {J.~E.}\ \bibnamefont
  {Marsden}}\ and\ \bibinfo {author} {\bibfnamefont {M.}~\bibnamefont {West}},\
  }\href@noop {} {\bibfield  {journal} {\bibinfo  {journal} {Acta Numer.}\
  }\textbf {\bibinfo {volume} {10}},\ \bibinfo {pages} {357} (\bibinfo {year}
  {2001})}\BibitemShut {NoStop}%
\bibitem [{\citenamefont {Qin}\ \emph {et~al.}(2009)\citenamefont {Qin},
  \citenamefont {Guan},\ and\ \citenamefont {Tang}}]{qin2009variational}%
  \BibitemOpen
  \bibfield  {author} {\bibinfo {author} {\bibfnamefont {H.}~\bibnamefont
  {Qin}}, \bibinfo {author} {\bibfnamefont {X.}~\bibnamefont {Guan}}, \ and\
  \bibinfo {author} {\bibfnamefont {W.~M.}\ \bibnamefont {Tang}},\ }\href@noop
  {} {\bibfield  {journal} {\bibinfo  {journal} {Physics of Plasmas
  (1994-present)}\ }\textbf {\bibinfo {volume} {16}},\ \bibinfo {pages}
  {042510} (\bibinfo {year} {2009})}\BibitemShut {NoStop}%
\bibitem [{\citenamefont {Mur}(1981)}]{mur1981absorbing}%
  \BibitemOpen
  \bibfield  {author} {\bibinfo {author} {\bibfnamefont {G.}~\bibnamefont
  {Mur}},\ }\href@noop {} {\bibfield  {journal} {\bibinfo  {journal} {IEEE T.
  Electromagn. C.}\ ,\ \bibinfo {pages} {377}} (\bibinfo {year}
  {1981})}\BibitemShut {NoStop}%
\bibitem [{\citenamefont {Maingi}\ \emph {et~al.}(2003)\citenamefont {Maingi},
  \citenamefont {Bell}, \citenamefont {Bell}, \citenamefont {Bush},
  \citenamefont {Fredrickson}, \citenamefont {Gates}, \citenamefont {Gray},
  \citenamefont {Johnson}, \citenamefont {Kaita}, \citenamefont {Kaye} \emph
  {et~al.}}]{maingi2003h}%
  \BibitemOpen
  \bibfield  {author} {\bibinfo {author} {\bibfnamefont {R.}~\bibnamefont
  {Maingi}}, \bibinfo {author} {\bibfnamefont {M.}~\bibnamefont {Bell}},
  \bibinfo {author} {\bibfnamefont {R.}~\bibnamefont {Bell}}, \bibinfo {author}
  {\bibfnamefont {C.}~\bibnamefont {Bush}}, \bibinfo {author} {\bibfnamefont
  {E.}~\bibnamefont {Fredrickson}}, \bibinfo {author} {\bibfnamefont
  {D.}~\bibnamefont {Gates}}, \bibinfo {author} {\bibfnamefont
  {T.}~\bibnamefont {Gray}}, \bibinfo {author} {\bibfnamefont {D.}~\bibnamefont
  {Johnson}}, \bibinfo {author} {\bibfnamefont {R.}~\bibnamefont {Kaita}},
  \bibinfo {author} {\bibfnamefont {S.}~\bibnamefont {Kaye}},  \emph {et~al.},\
  }\href@noop {} {\bibfield  {journal} {\bibinfo  {journal} {Nucl. Fusion}\
  }\textbf {\bibinfo {volume} {43}},\ \bibinfo {pages} {969} (\bibinfo {year}
  {2003})}\BibitemShut {NoStop}%
\bibitem [{\citenamefont {Strait}\ \emph {et~al.}(1995)\citenamefont {Strait},
  \citenamefont {Lao}, \citenamefont {Mauel}, \citenamefont {Rice},
  \citenamefont {Taylor}, \citenamefont {Burrell}, \citenamefont {Chu},
  \citenamefont {Lazarus}, \citenamefont {Osborne}, \citenamefont {Thompson},\
  and\ \citenamefont {Turnbull}}]{PhysRevLett.75.4421}%
  \BibitemOpen
  \bibfield  {author} {\bibinfo {author} {\bibfnamefont {E.~J.}\ \bibnamefont
  {Strait}}, \bibinfo {author} {\bibfnamefont {L.~L.}\ \bibnamefont {Lao}},
  \bibinfo {author} {\bibfnamefont {M.~E.}\ \bibnamefont {Mauel}}, \bibinfo
  {author} {\bibfnamefont {B.~W.}\ \bibnamefont {Rice}}, \bibinfo {author}
  {\bibfnamefont {T.~S.}\ \bibnamefont {Taylor}}, \bibinfo {author}
  {\bibfnamefont {K.~H.}\ \bibnamefont {Burrell}}, \bibinfo {author}
  {\bibfnamefont {M.~S.}\ \bibnamefont {Chu}}, \bibinfo {author} {\bibfnamefont
  {E.~A.}\ \bibnamefont {Lazarus}}, \bibinfo {author} {\bibfnamefont {T.~H.}\
  \bibnamefont {Osborne}}, \bibinfo {author} {\bibfnamefont {S.~J.}\
  \bibnamefont {Thompson}}, \ and\ \bibinfo {author} {\bibfnamefont {A.~D.}\
  \bibnamefont {Turnbull}},\ }\href {\doibase 10.1103/PhysRevLett.75.4421}
  {\bibfield  {journal} {\bibinfo  {journal} {Phys. Rev. Lett.}\ }\textbf
  {\bibinfo {volume} {75}},\ \bibinfo {pages} {4421} (\bibinfo {year}
  {1995})}\BibitemShut {NoStop}%
\bibitem [{\citenamefont {Porkolab}\ and\ \citenamefont
  {Chang}(1972)}]{porkolab2003instabilities}%
  \BibitemOpen
  \bibfield  {author} {\bibinfo {author} {\bibfnamefont {M.}~\bibnamefont
  {Porkolab}}\ and\ \bibinfo {author} {\bibfnamefont {R.}~\bibnamefont
  {Chang}},\ }\href@noop {} {\bibfield  {journal} {\bibinfo  {journal} {Phys.
  Fluids}\ }\textbf {\bibinfo {volume} {15}},\ \bibinfo {pages} {283} (\bibinfo
  {year} {1972})}\BibitemShut {NoStop}%
\bibitem [{\citenamefont {Sugaya}(1989)}]{sugaya1989nonlinear}%
  \BibitemOpen
  \bibfield  {author} {\bibinfo {author} {\bibfnamefont {R.}~\bibnamefont
  {Sugaya}},\ }\href@noop {} {\bibfield  {journal} {\bibinfo  {journal} {J.
  Phys. Soc. Jpn.}\ }\textbf {\bibinfo {volume} {58}},\ \bibinfo {pages} {1611}
  (\bibinfo {year} {1989})}\BibitemShut {NoStop}%
\end{thebibliography}%

\end{document}